\documentstyle[epsfig,subfigure]{mn}
 at 10truept
\title
     [Nonlinear Parameter Information]
{\vglue-3.0truecm
\centerline{\it\small For submission to Monthly Notices}
\vglue 2.5truecm
	Parameter information from nonlinear cosmological fields
\author
     [A.N. Taylor \& P.I.R. Watts]
     {A.N. Taylor  \& P.I.R. Watts\\
     Institute for Astronomy, 
     University of Edinburgh,
     Royal Observatory,
     Blackford Hill, 
     Edinburgh, 
     U.K.\\
	ant@roe.ac.uk, pirw@roe.ac.uk}}
\def\bib{\parskip=0pt\par\noindent\hangindent\parindent
    \parskip =2ex plus .5ex minus .1ex}

\newcommand{\be}{\begin{equation}}
\newcommand{\ee}{\end{equation}}
\newcommand{\ba}{\begin{eqnarray}}
\newcommand{\ea}{\end{eqnarray}}

\newcommand{\nn}{\nonumber \\}
\newcommand{\nnb}{\begin{displaymath}}
\newcommand{\nne}{\end{displaymath}}
\newcommand{\de}{\partial}

\newcommand{\x}{\mbox{\boldmath $x$}}

\newcommand{\k}{\mbox{\boldmath $k$}}

\newcommand{\kh}{\hat{\k}}

\newcommand{\C}{\mbox{\boldmath $C$}}

\newcommand{\J}{\mbox{\boldmath $J$}}
\newcommand{\T}{\mbox{\boldmath $T$}}
\newcommand{\khatb}{\mbox{\boldmath $\hat{ k}$}}
\newcommand{\rhatb}{\mbox{\boldmath $\hat{ r}$}}

\newcommand{\eF}{{\cal F}}
\newcommand{\0}{\mbox{\boldmath $0$}}
\newcommand{\F}{\mbox{\boldmath $\eF$}}
\newcommand{\thetab}{\mbox{\boldmath $\theta$}}

\newcommand{\phib}{\mbox{\boldmath $\phi$}}	
\newcommand{\eL}{{\cal L}}
\newcommand{\Tr}{{\rm Tr}}

\newcommand{\Mpch}{\,h{\rm Mpc}^{-1}}

\newcommand{\rgl}{\rangle}
\newcommand{\lgl}{\langle}

\def\bib{\parskip=0pt\par\noindent\hangindent\parindent
    \parskip =2ex plus .5ex minus .1ex}

\begin{document}

\maketitle

\begin{abstract}
We develop a general formalism for analysing parameter information
from non-Gaussian cosmic fields.  The method can be adapted to include
the nonlinear effects in galaxy redshift surveys, weak lensing surveys
and cosmic velocity field surveys as part of parameter estimation. It
can also be used as a test of non-Gaussianity of the Cosmic Microwave
Background.  Generalising Maximum Likelihood analysis to second-order,
we calculate the nonlinear Fisher Information matrix and likelihood
surfaces in parameter space.  To this order we find that the
information content is always increased by including nonlinearity.
Our methods are applied to a realistic model of a galaxy redshift
survey, including nonlinear evolution, galaxy bias, shot-noise and
redshift-space distortions to second-order.  We find that including
nonlinearities allows all of the degeneracies between parameters to
be lifted. Marginalised parameter uncertainties of a few percent
will then be obtainable using forthcoming galaxy redshift surveys.

%We find that, to second-order, all of the degeneracies between
%parameters are lifted allowing maginalised parameter uncertainties of a
%few percent to be obtained with
%The marginalised parameter
%uncertainties are improved by up to an order of magnitude. In
%addition, to second-order all of the degeneracies between parameters
%are lifted.
\end{abstract}

\section{Introduction}

%Generalities.
The estimation of cosmological parameters from galaxy redshift
surveys, Cosmic Microwave Background (CMB) surveys, weak lensing
surveys and velocity field surveys has become a major industry in
cosmology, due to the data explosion the field is currently
undergoing. To date the majority of the analysis has considered the
problem of parameter estimation from the large-scale, linear
regime. However much of the important information on parameters lies
on smaller scales, where nonlinear gravitational clustering has
distorted the initial field.  In the case of galaxy redshift surveys
additional nonlinearity may arise form the bias relationship between
galaxies and matter, while in the case of the CMB the original pattern
is still evident, allowing one to test the temperature fluctuations
%pattern
for evidence of primordial non-Gaussianity.

% Previous work.
% Linear likelihoods.
The industry standard for parameter estimation in the linear regime is
the multivariate Gaussian Likelihood method. Within this framework the
uncertainty on parameters can be estimated from the second derivatives
of the likelihood function, the Fisher Information matrix.  Tegmark,
Taylor \& Heavens (1997) were the first to study the Gaussian Fisher
Information matrix as a general tool for characterising the
information content of cosmological surveys, while Heavens \& Taylor
(1997) demonstrated its importance for the design of galaxy redshift
surveys. Tegmark (1997) used the Gaussian Fisher matrix to
investigate the accuracy of parameter estimation from the Sloan
Digital Sky Survey 
%galaxy redshift surveys 
in some detail, but did not calculate the effects of nonlinear
clustering, redshift-space distortions, or nonlinear biasing. Jungman
et al. (1996) applied a Fisher analysis to the CMB, while Hu \&
Tegmark (1999) used it to estimate parameter uncertainties in weak lensing
surveys.

% Nonlinear likelihoods.
While likelihood parameter estimation in the linear, Gaussian regime
has been extensively studied, until recently little work has been done
on the non-Gaussian generalisation. Amendola (1996) and Rocha et
al. (2000) have extended the likelihood methods to include
non-Gaussian contributions by expanding the likelihood function in a
series of cumulants. Amendola considered the effects of degradation of
parameter uncertainties due to the non-Gaussian broadening of the
likelihood function, while Rocha et al. applied their likelihood to
estimate residual non-Gaussianity in the CMB.

In the analysis of galaxy redshift surveys, Matarrese Verde \& Heavens
(1997) have investigated a Gaussian likelihood estimator for the
bispectrum as a tool for extracting the mean density parameter and
galaxy bias from galaxy redshift surveys in the nonlinear regime. This
builds on the work of Fry (1994) who suggested that the bispectrum
could be used to break the linear degeneracy between the bias
parameter and the density parameter, $\Omega_m$.  Verde et al. (1998)
developed this further to include redshift-space distortions.
Scoccimarro et al. (1998) and Scoccimarro, Couchman \& Frieman (1999)
similarly investigated the bispectrum, including the effects of
redshift-space distortions and galaxy bias.  Scoccimarro et al. (2000)
have recently estimated the bias parameters from the combined
IRAS-QDOT and 2Jy and 1.2Jy redshift surveys.

In this paper we develop a new method for exploring the estimation of
cosmological parameters from scales going into the nonlinear
regime. We expand the general non-Gaussian multivariate Probability
Distribution Function (PDF) and derive the non-Gaussian Fisher
matrix. Since the most relevant non-Gaussianity will be due to
nonlinear evolution of the density field, we use second-order
perturbation theory to find the dependance of the higher-order moments
on cosmological parameters and estimate the Fisher matrix to second
order.  We illustrate these methods using a realistic model galaxy
redshift survey, but note that they can be similarly applied to a  2-d weak
lensing, or 3-d cosmic velocity survey.

The paper is organised as follows. In Section 2 we describe the
expansion of the multivariate probability distribution function in
both the discrete and continuum limits. In Section 3 we develop the
formalism for calculating the Fisher information matrix in the
non-Gaussian regime.  We introduce a new function, the parameter
entropy, to study the likelihood surface in parameter space in Section
4 and calculate a general expression for non-Gaussian fields. These
methods are illustrated in Section 5 where we apply them to a model
galaxy redshift survey, including the effects of shot-noise,
redshift-space distortions and galaxy bias. We show how nonlinear
effects change the estimation of parameters in two surveys;  
%from surveys such as the
%Point Source Catalogue redshift survey (PSCz; Saunders et al. 2000),
the 2-degree Field galaxy redshift survey (2dF; Colless 1996) and the
Sloan Digital Sky Survey (SDSS; Gunn 1995).  Finally, we
present our conclusions in Section 5. Three appendices deal with
technical issues.

\section{Expansion of the multivariate PDF}
    
\subsection{The discrete distribution}

Cosmological data sets can be modeled by a random field, $\phi(\x)$, that is 
completely defined
by its multivariate Probability Distribution Function (PDF),
$\rho[\phi(\x)]$.  The field $\phi(\x)$ can be discretised and treated like an
$N$-dimensional data vector, $\phib$, whose components, $\phi_i$, may
represent, for example, an array of pixel values or the
amplitude of a set of harmonic modes. The PDF can be written as the
Fourier transform of a characteristic function;
\be
    \rho[\phib] = \int \!D[\J]\, G[\J] e^{i \J.\phib}
    \label{equ1}
\ee
where $\int \!D[\J]=\int\!\prod d^N\! J/(2\pi)^N$ is a multidimensional 
integral.
The characteristic function can be expressed by a set of
cumulants, defined by the series
\be
\ln G[\J] = \sum_n \frac{i^n}{n!}
\lgl \phi_{j_1}...\phi_{j_n} \rgl_c J_{j_1}...J_{j_n},
\ee
where there is an implicit summation over the $j_n$, running from $1$ to $N$. 
The $\lgl \phi_{j_1} \cdots \phi_{j_n} \rgl_c$ are the $n^{th}$ order
cumulants, or connected moments, of the field.

If only the second cumulant exists the distribution is a Gaussian.  If
the initial distribution is Gaussian we can separate out the $n=2$
term and then deal with the rest of the terms in the series.
Expanding out this second exponential to first order, keeping only
terms to $n=3$ and substituting back into equation (\ref{equ1}) we find

\ba
   	\rho[\phib] &=& \int \! D[\J] \,\exp{\big[-\frac{1}{2}J_i C_{ij} 
	J_j\big]}\nn
	& & \times \,  \left(1-\frac{i}{3!} D_{ijk} J_i J_j J_k \right)
         	\exp{(i J_m \phi_m)}\nn
          &=& \int \!D[\J] \,\exp{\big[-\frac{1}{2}J_i C_{ij} J_j\big]} \nn
          & & \times \,  \left(1-\frac{1}{3!} D_{ijk} \frac{\de^3}
		{\de \phi_i \de \phi_j \de \phi_k} \right)
            	\exp{(i J_m \phi_m)}          
\label{approxexp}            
\ea
where $C_{ij}=\lgl \phi_i \phi_j \rgl_c$ is the data covariance matrix and
$D_{ijk} = \lgl \phi_i \phi_j \phi_k \rgl_c$ is the data skewness. In the
second line of equation (\ref{equ2}) we have used the substitution $i \J =
\de/\de {\phib}$. Evaluating the Gaussian integral gives
\be
     \rho[\phib] = 
	\left(1- \frac{1}{3!} D_{ijk} \frac{\de^3}
		{\de \phi_i \de \phi_j \de \phi_k}\right)
	\frac{\exp{\big[- \frac{1}{2} \phi_i C^{-1}_{ij} \phi_j\big]}}{
		\sqrt{2 \pi {\rm det} C}}
\label{equ2}
\ee

This procedure is a multivariate generalisation of the Edgeworth
expansion for a single variable (Juszkiewicz et al. 1995, Bernardeau \&
Kofman 1995).  Finally, performing the differentiation we find
\ba
    \rho[\phib] &=& 
    \Big[1- \frac{1}{6} D_{ijk}\Big( 3  C^{-1}_{a[i} C^{-1}_{jk]}\phi_a
  - \, C^{-1}_{ia} C^{-1}_{jb} C^{-1}_{kb} \phi_a \phi_b \phi_c \Big)\Big] \nn
	&& \times \frac{\exp{\big[-\frac{1}{2} \phi_i C^{-1}_{ij} \phi_j\big]}}
	{\sqrt{2 \pi {\rm det} C}},
\label{edge}
\ea where $A_{[i_1 \cdots i_n]} \equiv 1/n! (A_{i_1 \cdots i_n} + {\rm
perms})$ is the symmetrization operator. 
This non-Gaussian multivariate distribution is the first major result of this paper.

%
%The advantage of
%using equation (\ref{edge}) over a Gaussian likelihood is its
%correction for nonlinear effects. In addition, in Gaussian likelihoods
%for higher moments, the variance of the moment being estimated must
%also be calculated. This means that if one is estimating parameters
%from $\lgl \phi^n \rgl $, the likelihood function would also require
%$\lgl \phi^{2n} \rgl $ to be calculated.

A well-known problem with truncating the moments, as we have done,
is that the resulting distribution is not a well-defined probability
distribution function.  In fact the problem lies not in the truncation
of moments, as the characteristic function is still well-defined, but
in the approximate calculation of the PDF itself in equation
(\ref{approxexp}). Hence the Edgeworth series, although properly
normalised, can produce non-physical oscillations and negative values
if the variance and skewness are pushed too far. This is a breakdown
of the series approximation, and one should take care not to force the
PDF into this regime.

Equation (\ref{edge}) can be generalised to arbitrary order. The input
for the nonlinear multivariate PDF are the higher order moments of the field.
These can either be left arbitrary and measured from the data
using the PDF as a likelihood function, or estimated by
perturbation theory and used to estimate other parameters from the
Bayesian relation
\be	
	\eL(\thetab|\x) = \rho(\thetab) \rho(\x|\thetab).
\ee
where we have defined $\thetab = (\theta_1,\theta_2,...,\theta_m)$ as the
$m$-dimensional parameter vector. Usually we will assume a uniform
prior, $\rho(\thetab)={\rm const}$.

%and $\int \!D[\J]=\int\!\prod d^N\! \J/[2 \pi]^N$ is again a multidimensional 
%integral, $C=\lgl \x^t \x \rgl_c$ is the data covariance matrix and
%$B_{ijk} = \lgl x_i x_j x_k \rgl_c$ is the data skewness. In the
%second line of equation (\ref{equ2}) we have used the substitution $i \J =
%\de/\de \x$, 
%and in the last line inverse Fourier transformed the
%Gaussian integral. This procedure is a multivariate
%generalisation of the Edgeworth expansion for a single variable.
%Finally, performing the differentiation we find
%\ba
%    \rho[\x] &=& \frac{e^{-\frac{1}{2} \x^t \C^{-1} \x}}{\sqrt{2 \pi {\rm det} C}}
%    \Big[1- \frac{1}{3!} \B(\C^{-1}\!\x\, \C^{-1} +\C^{-1}\!\x\, \C^{-1} \nn
%	& & + \C^{-1} \C^{-1}\! \x 
%        -
%	\C^{-1}\!\x\, \C^{-1}\!\x\, \C^{-1}\!\x) \Big]
%\label{edge}
%\ea
%where the first three terms in the inner bracket permutate integers.

In developing the methodology it will be useful to express the
perturbed Gaussian distribution as 
\be 
	\rho[\phib] = \rho_0[\phib](1-X)
\label{perturb}
\ee 
where $\rho_0[\phib]$ is the underlying Gaussian distribution and $X$
is a small perturbation. It is worth noting that some of the results
of this paper are quite general and apply to all cases where the
multivariate PDF is expressible in this way, for example the harmonic
oscillator approach of \mbox{Rocha et al. (2000)}, or multivariate generalisations
of other non-Gaussian fields.

\subsection{The continuum distribution}

In most calculations it is easier to consider the continuum distribution
of the function $\phi(\x)$.
This can be found by taking the limit $N \to \infty$ and substituting 
$\sum \rightarrow V \int d^3 k/(2\pi)^3$. From here on we choose to use a 
field embedded in 3 spatial dimensions, but the method is easily generalised
to an arbitrary number of dimensions. 
In the continuum limit the underlying Gaussian distribution is 
\be
	\rho_0[\phi] = \frac{1}{A^{1/2}} \exp{\left[-\frac{1}{2} 
	\int \! d^3 \!x_1 d^3\! x_2\, 
	\phi(\x_1) C^{-1}(\x_1,\x_2) \phi(\x_2) \right]}
\label{gausscon}
\ee
where $A$ is a normalisation factor. 
The expression for X, in the continuum case, containing the higher-order 
moments of the field $\phi(\x)$,
is given in Appendix A. 
The covariance and bivariance functions are, respectively 
\be
C(\x_1,\x_2) = \langle \phi(\x_1)\phi (\x_2)\rangle_c
\ee
\be
D(\x_1,\x_2,\x_3)  = \langle \phi(\x_1)\phi(\x_2)\phi(\x_3)\rangle_c
\ee
and the inverse function $C^{-1}(\x_1,\x_2)$ is defined by
\be
\int \! d^3 \! x \, C^{-1}(\x_1,\x) C(\x,\x_2) = \delta_D(\x_1 - \x_2).
\ee

\section{The non-Gaussian Fisher matrix}
\label{fisher}
To examine the information content available in nonlinear data it
is useful to construct the Fisher information matrix for
parameters. One issue of interest is the flow of information about
parameter values during nonlinear evolution. In one respect nonlinear
evolution destroys information about initial conditions. But since
information has to be preserved, it must inevitably be transported
up the hierarchy of correlation functions. However nonlinear growth
itself depends on the detailed values of cosmological parameters.  The
question is does nonlinear evolution create or destroy parameter
information?

To explore these questions we can define three useful quantities:
the entropy of the system,
\be
	S \equiv \lgl \eL \rgl = - \int \! d\rho \, \ln \rho,
\ee
where $\eL= -\ln L$ is the average log--likelihood evaluated at the 
true parameter values;
the average gradient of the log likelihood in parameter space,
\be
	\lgl \eL_i \rgl = - \int \! d \rho \, \de_i \ln \rho = 0,
\ee
where $\de_i$ is the gradient in the $i$ direction of parameter space, 
and is identically zero at the maximum likelihood values; and
the Fisher Information matrix, the curvature of the likelihood surface about its
maximum, which
quantifies our knowledge of a set of parameters,
\be
	\eF_{ij} \equiv \lgl \eL_{ij} \rgl = - \int \! d \rho \,
	\de_i\de_j\ln\rho = 
	\int \! d \rho \,  (\de_i \ln \rho) (\de_j \ln \rho).
\ee
The parameter covariance matrix can be constructed from the Fisher matrix
via
\be
    \T \equiv \lgl \delta \thetab^t \delta \thetab \rgl = \F^{-1},
\ee
where $\delta \thetab$ is the displacement from the maximum likelihood
point, $\delta \thetab = \thetab - \thetab'$.

\subsection{Two-point correlations}

For the case of a Gaussian PDF we may derive
a simple and familiar result for the Fisher information matrix. Since
all higher order correlations are zero, $X = 0$. The Fisher
matrix for Gaussian fields is therefore  
\be
\eF_{0,ij} = - \int\! d\rho_0 \, \de_i \de_j  \ln{\rho_0}.
\label{fish1}
\ee
 The calculation is simplified if we  
choose the random field
$\phi(\x)$, to be the set of Fourier modes, $\delta(\k)$, where the
covariance matrix is diagonal due to spatial invariance;
\ba 
	C(\k_1,\k_2) & = & \langle \delta(\k_1)\delta(\k_2) \rangle \nn
 	& = & (2\pi)^3 P(k) \delta_D(\k_1 + \k_2).
\label{pow}
\ea
In this case the log-likelihood becomes
\be
	2 \eL_0 
	= \int \! \frac{d^3 \! k}{(2 \pi)^3}\, \frac{\delta(\k) \delta^*(\k)}{
		P(k)} + V\int \! \frac{d^3 \! k}{(2 \pi)^3}  \, \ln P(k)
\label{2ptlike}
\ee
where the second term is the normalisation factor modulo an unimportant 
constant term. Here we have used
the matrix relation $\ln{\det{\C}} = \Tr\ln{\C}$ before transforming to 
the continuum limit. For a finite volume survey we have also used the
approximation $\delta_D(\0)=V$, where $V$ is the survey volume. For a Gaussian
distribution the Fisher matrix, equation (\ref{fish1}), becomes
\ba
\eF_{0,ij} &=& \frac{V}{2}\int \! \frac{d^3 \! k}{(2 \pi)^3}  \,  
	(\de_i \ln{P(k)})(\de_j \ln{P(k)}).
\label{fishG}
\ea 
A similar expression to equation (\ref{fishG}) was previously obtained by 
Tegmark (1997), using a different derivation.

\subsection{Higher-order correlations}

The non-Gaussian regime may be explored via the perturbed
multivariate PDF, equation (\ref{perturb}). 
The gradient of the log-likelihood, 
$\lgl \eL_i \rgl$, is again identically zero, while the Fisher matrix becomes
\ba
\eF_{ij}\!\!\!\!&=&\!\!\!\! - \int \! d \rho_0 \, (1-X)(\de_i \ln \rho_0) 
(\de_j \ln \rho_0) \nn
& &\!\!\!\!+ 2 \int \! d \rho_0 \, X_{(i} \de_{j)} \ln \rho_0  
 - \int \! d \rho_0 \, X_i X_j/(1-X).
\ea
If the initial PDF is Gaussian, $\ln \rho_0$ is an even function of variables
and $X$, given by equation (\ref{Xfact}), is odd. Expanding the last term
this expression reduces to
\be
	\eF_{ij} = \eF_{0,ij} + \lgl X_i X_j \rgl_G,
\label{nonlinF}
\ee
which is accurate up to $X^4$. The brackets, $\lgl \cdots \rgl_G$, 
denote averaging over the Gaussian distribution.

Again the analysis is simplified if we work in a Fourier representation,
assuming spatial invariance. In this case the 
bivariance is
\ba
	D(\k_1,\k_2,\k_3) & = &\! \! \! \lgl \delta(\k_1)\delta(\k_2)
	\delta(\k_3)\rgl_c \nn
 	& = & \! \! \! (2\pi)^6 B(\k_1,\k_2,\k_3)\delta_D(\k_1+\k_2+\k_3),
\label{bisp}
\ea
which selects only those combinations of wavevectors that form closed
triangles. We will henceforth refer to $B(\k_1,\k_2,\k_3)$ as the
bispectrum.

With equations (\ref{pow}) and (\ref{bisp}) the perturbation term, $X$, 
reduces to
\be
	X = \frac{1}{6}\int \!\frac{d^3\!k_1}{(2\pi)^3} 
        \!\frac{d^3\!k_2}{(2\pi)^3} \,
	Q(\k_1,\k_2) \delta(\k_1)\delta(\k_2)\delta(-\k_1-\k_2),
\label{xconst}
\ee
where
\be
	Q(\k_1,\k_2) = \frac{B(\k_1,\k_2,-\k_1-\k_2)}{P(k_1)P(k_2)P(|\k_1+\k_2|)}.
\label{Qdeff}
\ee
This term derives from the $\phi^3$-order term in the perturbation of
the general PDF. The $\phi$-order term vanishes under spatial
invariance, since the bispectrum vanishes faster than the variance on
large scales. This final constraint will always be true for a
realistic PDF since the Jensen inequalities (Kendall \& Stuart, 1969)
state that
\be
	\lgl x^a \rgl <  \lgl y^b\rgl^{a/b},
\ee
where $x$ is an arbitrary random variables and $a>b$ are real numbers.
This implies that
\be
	\frac{\lgl \delta^3 \rgl}{\lgl \delta^2 \rgl} <
	\lgl \delta^2\rgl^{1/2},
\ee
so that the ratio of the bispectrum to the power will always vanish 
faster than the variance.

The log-likelihood in the nonlinear regime is
\be
	\eL =  \eL_0 +X +\frac{1}{2}X^2
\ee
where $\eL_0$ is the 2-point likelihood function, (\ref{2ptlike}),  and $X$ 
is given by (\ref{xconst}). This expression can then be used
to find the maximum likelihood values of parameters for systems in which 
the covariance matrix is close to diagonal.

We calculate the $\lgl X_i X_j \rgl_G$ term in equation
(\ref{nonlinF}) making use of the cumulant expansion theorem to expand
the 6-point correlation function into its connected parts. For the
case of a Gaussian average only the two point terms are non-zero so that
\be
	\lgl \delta_1 \delta_2 ...\delta_6 \rgl = 
\lgl \delta_1\delta_2 \rgl_c\lgl \delta_3\delta_4 \rgl_c\lgl
\delta_5\delta_6 \rgl_c + {\rm perms}.
\label{wicks}
\ee
Of these terms only six survive due to spatial invariance. 
%This simplification is a nice result that comes from our choice to work
%with the $\delta$-field explicitly rather than with it's
%moments. 
Equation (\ref{wicks}) essentially states that to the order we are
working to, the covariance of bispectrum triangles is zero.  Matarrese
et al. (1997), who use a Gaussian likelihood for the bispectrum,
calculate the covariance matrix of the bispectrum to higher order and
show that correlations between triangles appear through the pentaspectrum.

The non-Gaussian Fisher matrix can subsequently be written,
\ba
\eF_{ij} & = & \eF_{0,ij} + \frac{V}{6}\int \!\frac{d^3\!k_1}{(2\pi)^3} 
\! \frac{d^3 \!k_2}{(2\pi)^3} \, Q_i(\k_1,\k_2) Q_j(\k_1,\k_2) \nn
 &&  \times   P(k_1) P(k_2) P(|\k_1+\k_2|), 
\label{nongausfish}
\ea
where the subscripts $i$ and $j$ denote differentiation with 
respect to the parameters. This is the second major result of this paper,
and is independent of the nature of the non-Gaussianity, other than 
statistical spatial invariance of the field.

With equations (\ref{nongausfish}) and (\ref{nonlinF}) we can now
address the question of information flow. Since the diagonal terms
$\lgl X_i^2 \rgl_G$ are positive, the addition of the bispectrum term
can only add information. This is because the dominant effect, to this
order, is the additional information brought by the higher order
moments. Nonlinear evolution will also change the shape of the PDF and
degrade parameter information, but we see here that this is a
higher-order effect.

\section{The Parameter Entropy Function}
\label{secentropy}
The Fisher Information matrix provides a compact way of determining both
marginal and conditional errors on a given set of
parameters. However, in some circumstances it is useful to have more
information about the detailed shapes of likelihood surfaces in parameter
space. This may be achieved by generalising the definition of the
entropy from Section \ref{fisher}  so that 
\be
S(\theta) = \lgl \eL(\theta) \rgl = - \int d\rho' \ln{\rho},
\ee
where a prime denotes that a quantity is evaluated at its true (maximum
likelihood) value. With this function one can map out the
likelihood distribution of parameter space.

An advantage of the parameter entropy is that it can be
calculated directly from the covariance and bivariance functions, 
without having to take parameter derivatives. This may be a
noisy process if the spectra are generated numerically, ie 
directly from a Boltzmann solver. 
Noise in the Fisher matrix can spuriously break parameter
degeneracies, and the entropy is one way to avoid this. 
Noise may also prevent the Fisher matrix from being positive definite,
and hence unphysical.
Strong correlations between parameters can also make
the Fisher matrix ill-conditioned and numerically difficult to invert.
The entropy clearly contains more
information about the parameters and the shape of likelihood contours,
which may be important for investigating degeneracies among parameters.
Finally, the parameter entropy does not assume the parameter
likelihood distribution is Gaussian, which is implicit in the Fisher
matrix.  A  disadvantage of the entropy is that evaluating the
marginal parameter uncertainties can be difficult since the full
distribution of $S$ must be mapped out around its maximum in parameter
space. 

%The Fisher Information matrix provides a compact method of determining
%the information content of a data set. However, in some instances more
%information is required about the shape of likelihood surfaces in
%parameter space. One way to quantify this information is by introducing
%a parameter entropy function, in the sense of Shannon
%\be
%S(\theta) = \lgl \eL(\theta) \rgl = \int d\rho' \ln{\rho}
%\ee
In the Gaussian case the entropy is given by
\be
S_0 = - \int\! d \rho'_0  \,\ln{\rho_0} .
\ee
In terms of the power spectrum this gives, up to an unimportant constant,
\be
	S_0 = \frac{V}{2} \int\! \frac{k^2 dk }{2 \pi^2} 
	\left(\frac{P'(k)}{P(k)} +  \ln{P(k)}\right). 
\label{entlin}
\ee
The nonlinear evolution of this quantity follows from the results of
the last section. Expanding the non-Gaussian PDF as in equation
(\ref{perturb}) gives
\be
	S= S_0 + \int \! d \rho'_0 \,  X' \ln \rho_0  - \int \! d \rho'_0 \,
  (1-X') \ln (1-X) 
\label{ent1} 
\ee
Again if $\rho_0$ is a Gaussian the second term in equation (\ref{ent1})
vanishes. The final log term can be expanded to second order in $X$, $\ln(1-X)
\approx -X - X^2/2$, yielding
\be
	S = S_0 - \lgl X X' \rgl_G + \frac{1}{2} \lgl X^2 \rgl_G ,
\label{entnon}
\ee
This expression again simplifies using a Fourier representation and, with 
the results from the previous section, we find
\ba
	\lgl X^2\rgl_G \!\!\! &=& \!\!\!
	\frac{V}{6}\int \! \frac{d^3 \!k_1}{(2\pi)^3} 
	 \frac{d^3 \!k_2}{(2\pi)^3}\, 
	Q^2(\k_1,\k_2)  P'(k_1) P'(k_2) \nn
	&& \times
 	   P'(|\k_1+\k_2|) 
\label{entfirst}
\ea
and
\be
	\lgl X X' \rgl_G  =  \frac{V}{6}\int\! \frac{d^3 \!k_1}{(2\pi)^3} 
	 \! \frac{d^3 \!k_2}{(2\pi)^3}\, Q(\k_1,\k_2) 
 	 B'(\k_1,\k_2,-\k_1-\k_2).
\label{entsecond}
\ee
Equations (\ref{entnon}), (\ref{entfirst}) and (\ref{entsecond}) are the third major 
new result of this paper.

The parameter entropy evaluated at the maximum likelihood point,
tells us about the order in the system and can be written
\be
	S= S_0 - \frac{1}{2} \lgl X^2 \rgl_G,
\label{entropy}
\ee
where 
\be
	S_0 = \frac{V}{2} \int\! \frac{k^2 dk }{2 \pi^2} \ln P(k)
\ee
again up to an unimportant constant. In the case of gravitational
instability, the power spectrum will always increase with time, so the
linear entropy, $S_0$, will also increase. As the field is Gaussian
distributed, this corresponds to the maximum amount of disorder
possible -- that is the Gaussian requires the least information to
specify it.

The effect of the non-Gaussian term in equation (\ref{entropy}) always
acts to {\em decrease} the entropy of the system, relative to the 
Gaussian case. While this may seem
counter to our usual expectation that systems evolve towards higher
entropy states, the effect of non-Gaussianity here introduces order,
or structure, into the system, reducing the entropy.

%\begin{figure}
%\centering
%\begin{picture}(200,200)
%\special{psfile='pow.ps' angle=0 voffset=-100 hoffset=-30 vscale=50 hscale=50}
%\end{picture}
%\caption{The matter power-spectrum in real space, and the galaxy  power-spectrum 
%in redshift space.
%}
%\label{fig1}
%\end{figure}

%\begin{figure}
%\centering
%\begin{picture}(200,200)
%\special{psfile='bi.ps' angle=0 voffset=-100 hoffset=-30 vscale=50 hscale=50}
%\end{picture}
%\caption{The matter  bispectrum for equilateral triangles in
%real space, and the galaxy   bispectrum for equilateral triangles
%in redshift space.
%}
%\label{fig1}
%\end{figure}

\section{Application to Galaxy Redshift Surveys}
\label{grs}

To illustrate the effects of including higher order statistics in the
likelihood function it is useful to calculate the Fisher matrix and
entropy for a simple analytic model. Since much can be derived from
the nonlinear regime of galaxy redshift surveys, we choose this as our
model, with application to the 2dF and Sloan Digital Sky Survey.
We include sufficient detail in the model so that it possesses
most of the properties associated with galaxy redshift surveys.  The
random field of interest is therefore the biased, redshift space
distorted mass density contrast $\delta^s = \delta n^s/n^s$. 

Again, for simplicity, we elect to work in Fourier space.  On large
scales it is both more efficient and more accurate to exploit the
natural symmetry of the survey and work in spherical harmonics
(Heavens \& Taylor 1995, Tadros et al. 1999, Taylor et
al. 2000). While this is important for linear analysis, it is not
necessarily so for the nonlinear analysis. Nonlinear effects dominate
on smaller scales where the Fourier representation can replace
spherical harmonics to a high degree of accuracy. However, this
simplification may begin to break down for the case where triangles of
the bispectrum lead to the mixing of information from both large and
small scales (see section \ref{results}).

\subsection{Gravitational instability}
\label{gravinst}

For a galaxy redshift survey a major source of non-Gaussianity comes
from nonlinear evolution due to gravitational instability.  For
Gaussian initial conditions gravitational evolution may be modelled
via perturbation theory.  The skewness of the mass density field in
perturbation theory is given by (Goroff et al. 1986)
\ba
	B(\k_1,\k_2,\k_3) \!\!&=&\!\! (2 \pi)^3 [ 2 J(\k_1,\k_2) P(k_1)P(k_2)
	+ {\rm cyc}(23,13)]\nn & & \times \delta_D(\k_1+\k_2+\k_3), 
\ea 
where
\be 
	J(\k_1,\k_2) = 1 + \frac{1}{2} \mu_{12} \left( \frac{k_1^2+k_2^2
	}{ k_1 k_2}\right) + \kappa(\Omega_m) [\mu_{12}^2-1], 
\label{jdeff}
\ee
and where $\mu_{12}=\kh_1.\kh_2$. The dependance of $J(\k_1,\k_2)$ on
the density parameter is weak and comes in only through the function
$\kappa(\Omega_m)=D_2/D_1^2 \approx - 3/7 \, \Omega_m^{-2/63}$, the ratio of 
first and second order growth
functions (Peebles 1980, Bouchet et al. 1992). For simplicity we use
the Einstein--de-Sitter value of $\kappa = -3/7$ (Bouchet et al. 1992). 

\begin{figure}
\centering
\subfigure{\epsfig{figure=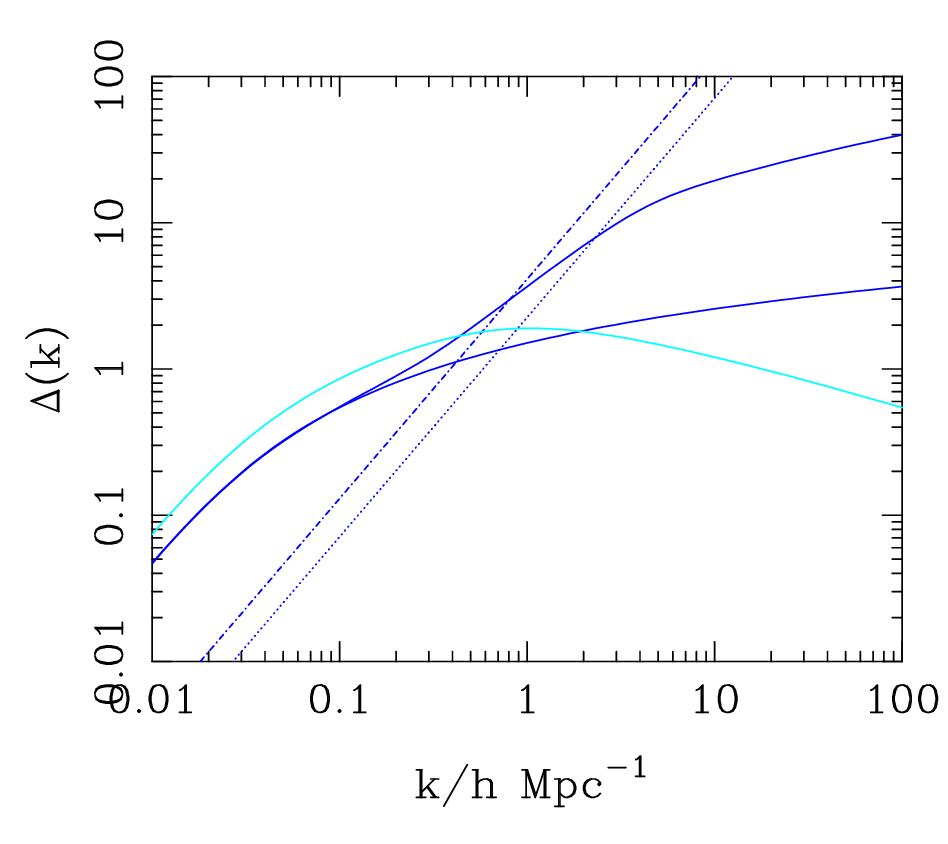,width=8.5cm,angle=0,clip=}}
\caption{Fiducial matter power spectrum for a model galaxy redshift
survey. The linear and nonlinear power in real-space (solid lines) are
plotted alongside the angle averaged redshift power spectrum (light
solid line) with small-scale damping included. The shot-noise per mode
for both of the model surveys are also plotted (see table
\ref{table1}; 2dF (dot-dash) and SDSS (dotted).}
%PSCz (dashed), 2dF (dot-dash) and SDSS (dotted)}
\label{figa}
\end{figure}

The CDM-type model power spectrum we shall use takes the form 
\be
	\Delta^2(k) = \delta_H^2 (k/H)^4 k^{n-1} T^2(k,h,\Omega_m), 
\ee 
where
\be 
	\Delta^2(k)\equiv k^3 P(k)/2 \pi^2
\ee 
and $T(k,h,\Omega_m)$ is the transfer function as given by Bardeen et
al. (1986).  The fiducial model we choose has parameters
$(\delta_H,h,\Omega_m,b_1,b_2)=(5\times
10^{-5},0.65,0.3,0.002,1,0.5)$. For simplicity we assume spatial
flatness. Figure \ref{figa} shows the spectrum used for subsequent
calculations.

\subsection{Choosing bispectrum triangles}
\label{triangles}
\begin{figure}
\centering
\subfigure{\epsfig{figure=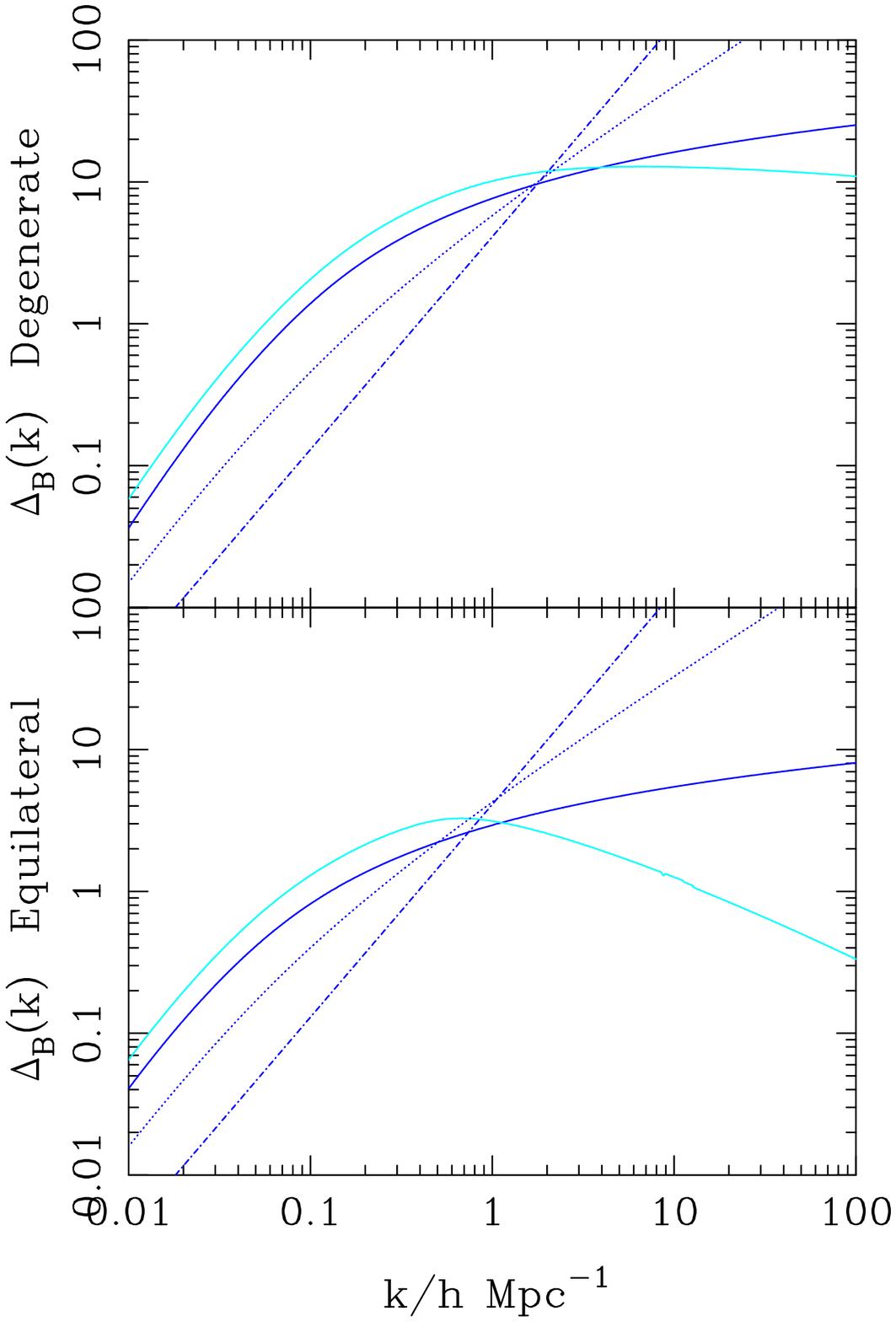,width=8.5cm,angle=0,clip=}}
\caption{Fiducial bispectra for the redshift survey model for
equilateral triangles (bottom panel) and degenerate triangles (top
panel). The real-space bispectra (heavy solid lines) are shown
alongside the redshift-space bispectra (light solid lines) with small
scale damping. The $P(k)/n$ noise term (dotted) and the $1/n^2$
noise term (dot-dashed) are also shown for a 2dF-type survey.}
\label{figb}
\end{figure}

While our approach so far has been to retain as much information as
possible, in practice it may not be possible to include all shapes of
triangles in $k$-space due to CPU restrictions. It therefore makes
sense to identify and study a subset of shapes that maximise the
information content. We discuss this issue in more detail in Section
\ref{results}. For simplicity we consider here only two distinct shapes:
equilateral triangles and degenerate triangles.  Degenerate triangles are
constructed from two parallel wavevectors plus a third that is
aligned anti-parallel with magnitude equal to the sum of the other two. A
convenient subset of the degenerate category, and the configuration we
shall consider in this analysis, has the magnitude of the two
smaller wavevectors being equal. Constraints are made by inserting an
appropriate dimensionless operator into equations
(\ref{nongausfish}) and (\ref{entnon}). For equilateral triangles we
use: 
\be 
	\Delta_{{\rm EQ}} = \int \! \frac{d^3\!k}{4\pi} \,
\delta_D(k_1-k) \delta_D(k_2-k)\delta_D(k_3-k); 
\ee
whereas for the subset of degenerate (DG) triangles one possible constraint is
\be
\Delta_{{\rm DG} } = \frac{1}{4}\int \frac{d\k}{4\pi} \delta_D(k_1-k)
\delta_D(k_2-k/2)\delta_D(k_3-k/2).
\ee
%Note that for a given angle between $k_1$ and $k_2$, the degenerate constraint
%may also be written with $k_1$ and $k_2$ interchanged. In the
%subsequent analysis we will always implicitly sum over both contributions. 
For the triangles of interest the bispectrum can be written
\be
	B(k) = \frac{12}{7} P^2(k) \;\;\;\;\;\; {\rm Equilateral}
\ee
\be
	B(k) = 4 P^2(k/2) - P(k/2) P(k) \;\;\;\;\;\;{\rm Degenerate}.
\ee
We plot the bispectra in Figure \ref{figb}, where we have defined 
the function 
\be
	\Delta^2_B(k) \equiv \frac{k^3\sqrt{B(k)}}{2 \pi^2}
\ee
which is useful to compare with $\Delta^2(k)$.

For an ideal, finite survey, in the absence of shot-noise and
redshift-space distortions, the nonlinear Fisher matrix for
equilateral triangles can conveniently be reduced to 
\ba
	\eF_{ij}& = & \frac{V}{2}\!\int \!\!\frac{k^2 d k}{2 \pi^2}
	\Big[1+\Big(\frac{12}{7}\Big)^2 \frac{\Delta^2(k)}{6}\Big] \nn
 & & \times \; [\de_i \ln{\Delta^2(k)}][\de_j \ln{\Delta^2(k)}].
\ea
Similar expressions for the parameter entropy and for the case of
degenerate triangles can be found in appendix B.

The effect of including nonlinear terms is two-fold. Firstly we
see that the bispectrum introduces a second, positive term into the 
Fisher matrix, driving the parameter uncertainty down. In addition, the
nonlinear correction to the likelihood allows one to push the wave-range
to higher $k$, again driving down the uncertainty.

%\begin{figure}
%\centering
%\begin{picture}(200,200)
%\special{psfile='derivP.ps' angle=0 voffset=-100 hoffset=-30 vscale=50 hscale=50}
%\end{picture}
%\caption{The derivatives of the matter power-spectrum in real space.
%}
%\label{fig1}
%\end{figure}

%\begin{figure}
%\centering
%\begin{picture}(200,200)
%\special{psfile='derivB.ps' angle=0 voffset=-100 hoffset=-30 vscale=50 hscale=50}
%\end{picture}
%\caption{The derivatives of the matter bispectrum in real space.
%}
%\label{fig1}
%\end{figure}%

%\begin{figure}
%\centering
%\begin{picture}(200,200)
%\special{psfile='para_nonlin_bi_red_comp.ps' angle=0 voffset=-100 hoffset=-30 vscale=50 hscale=50}
%\end{picture}
%\caption{The parameter uncertainty.
%The noise and
%volume are that of a 2df type survey.
%}
%\label{fig1}
%\end{figure}

%In the case of flat triangles the bispectrum reduces to
%\be
%	B(k,k/2,k/2) = P(k/2) [9P(k)+4 P(k/2)],
%\ee

\subsection{Poisson sampling of the density field}

Assuming the measured galaxy population can be modelled as a random
sampling of a smooth underlying field, the power- and bi-spectra
transform according to
\ba
	P(k) \!\!\! &\rightarrow& \!\!\!P(k) + \frac{1}{n}, \nn
	B(\k_1,\k_2,\k_3)\!\!\! &\rightarrow&\!\!\!
	 B(\k_1,\k_2,\k_3) \nn
	& & + \frac{1}{n}[P(k_1) + P(k_2) + P(k_3)]
	+ \frac{1}{n^2},
\ea
where $n$ is the mean number density of sources.  If the noise varies
across the survey, as is the case for flux-limited galaxy redshift
surveys, this can be included in the models by the substitution
$n=n(r)$ and $V\rightarrow \int \! d^3 \!r$ in the Fisher and entropy
estimates. For simplicity we work with the volume limited case and
consider two models characterised by the number density, $n$, and
survey volume, $V$, as given in \mbox{Table \ref{table1}}. Noise
levels for the surveys are plotted in \mbox{Figures \ref{figa} and
\ref{figb}.}

\begin{table}
\begin{center}
\begin{tabular}{ccc}\hline
 Survey & Mean Number & Effective \\
	& Density ($n$) & Volume (V) \\ \hline
% PSCz	&  $1\times10^{-3}$ & $2\times10^7$ \\
 2dF    &  $3\times10^{-3}$ & $9\times10^7$ \\	
 SDSS   &  $10\times10^{-3}$ & $10\times10^7$\\\hline
\end{tabular}
\caption{Parameters for two model redshift surveys considered in this paper.}
\label{table1}
\end{center}
\end{table}

\subsection{Bias and redshift space distortions}
\label{redshift}

%\begin{figure}
%\centering
%\subfigure{\epsfig{figure=fig1.eps,width=8.5cm,angle=0,clip=}}
%\caption{Parameter derivatives (top panel) and 1$\sigma$ error levels
%for the joint estimation of $b_1$ and $\eta$ in redshift space (bottom
%panel).  The top panel shows the angle averaged derivative of the $Q$
%function from equation (\ref{Qdeff}) for equilateral triangles (solid
%lines) and degenerate triangles (dotted lines) taken with respect to
%each of the parameters.  The bottom panel shows the joint errors on
%the parameters when using the Gaussian likelihood (solid line), when
%adding in the contribution from equilateral triangles (dashed line)
%and when adding in both equilateral and degenerate triangles (dotted
%line).  The survey parameters closely resemble those of the 2dF, as
%listed in table \ref{table1}. }
%\label{fig1}
%\end{figure}

The use of redshift coordinates as radial distance introduces
a distortion in the observed density field;
\be
	\delta^s(\k) = D(k\sigma \mu)(1+\beta \mu^2) \delta(\k),
\ee
where the linear distortion modulates the density by a factor
$(1+\beta \mu^2)$ (Kaiser 1987) with $\beta = \Omega_m^{0.6}/b_1$ and
$b_1$ the linear bias parameter. The small-scale pairwise velocity
dispersion can be modeled by convolution of the density field with a
random component along the line of sight (Peacock 1992, Peacock \&
Dodds 1994);
\be
	D(k\sigma \mu) = \frac{1}{\sqrt{1+k^2 \sigma^2 \mu^2/2}},
\ee
where $\mu=\rhatb.\khatb$.
Galaxy bias can be modeled by a local expansion of the density field
(Fry and Gaztanaga 1993), 
\be
	\delta_g = \sum_m \frac{b_m}{m!} \delta^m
\ee
where the coefficients $b_m$ are the bias parameters. The power
spectrum for a biased, redshifted, and noisy galaxy survey is therefore
\be
	P^s_g(\k) = b_1^2 D^2(k \sigma \mu)(1+\beta \mu^2)^2 P(k) + 
	\frac{1}{n}.
\label{redshiftpower}
\ee
Note that we only consider the case of a deterministic bias, but this
formalism can be easily generalised to include extra parameters for a
stochastic biasing scheme (Pen 1998, Dekel \& Lahav 1999).

For equilateral triangles the angles between the line of sight and the 
wavevectors of the bispectrum triangles are
\ba	
	\mu_1 &=& \mu, \nn
	\mu_2 &=& -\frac{1}{2} \mu + \frac{3}{2}\sqrt{1-\mu^2}, \nn
	\mu_3 &=& -\mu_1 -\mu_2,
\ea
where $\mu_i = \rhatb . \khatb_i$. The corresponding redshifted
bispectrum for equilateral triangles is (Hivon et al. 1995, Heavens et
al. 1998, Verde et al. 1998, Scoccimarro et al. 1999)
\ba
	B^s_g(\k) &=& D_2(k \sigma \mu)\Big[2 b_1^3 {\rm Ker}
	(k,\mu,\beta,b_1) \nn
	& &+ 3 b_2(1+\beta + 3/16 \beta^2)\Big] P^2(k)  \nn
	& &+ \frac{1}{n}  (D(k\mu_1) (1+\beta \mu_1^2)^2 + {\rm cyc})
	 P(k) + \frac{1}{n^2}
\label{bisp_eek}
\ea
where $b_2$ is the second-order bias term and where kernel ${\rm
Ker}(k,\mu,\beta,b_1)$ is provided in Appendix B. Similar
expressions for degenerate triangles can be obtained and are also
given in the appendix. The redshifted power and bi-spectra are shown
in Figures \ref{figa} and \ref{figb}.

\section{Results for a model galaxy redshift survey}
\label{results}

\subsection{Overview of parameter estimation}

In this Section we apply the model outlined in Section \ref{grs} to
estimate parameter uncertainties from two model galaxy redshift
surveys, based on the 2dF and SDSS (see Table 1 for survey
parameters). The parameters of the model are $(h,\delta_H, \Omega_m,
b_1, b_2)$, with fiducial values given in Section \ref{gravinst}. The
model enters the analysis via the redshifted galaxy power spectrum,
equation (\ref{redshiftpower}), and the bispectrum for equilateral
triangles, equation (\ref{bisp_eek}), and degenerate triangles,
equation (\ref{redshiftedbideg}). These are used to estimate the
function $Q(\k_1,\k_2,\k_3)$, given by equation (\ref{Qdeff}), which
is then used to calculate the nonlinear Fisher matrix, given by
equations (\ref{nongausfish}) and (\ref{fishG}). The parameter entropy
function is calculated from equations (\ref{entlin}), (\ref{entnon}),
(\ref{entfirst}) and (\ref{entsecond}).

This section is laid out as follows. In Section \ref{nonlinpara} we
present results for the marginalised parameter uncertainties for the
two surveys, as a function of the truncation wavenumber for the
Fourier integrals.  
Our basic result is that cosmological parameters
may be estimated to accuracies of a few percent using redshift surveys
independantly of other measurements.
%an order of magnitude
%improvement can be made in parameter estimation by including
%nonlinearity. 
In Section \ref{lifted} we show the correlation
coefficients for all the parameters, again as a function of
wavenumber, and illustrate that to second-order all of the
degeneracies between parameters are lifted.  We illustrate how
parameter estimation is improved using a simplified two-parameter
model in Section \ref{twopara}. We explore the relative information
content of equilateral and degenerate triangles in Section
\ref{triinfo}.  We begin with an estimation of the parameter
uncertainties from the full model.

\subsection{Nonlinear parameter estimation from redshift surveys}
\label{nonlinpara}

\begin{figure*}
\centering
\subfigure{\epsfig{figure=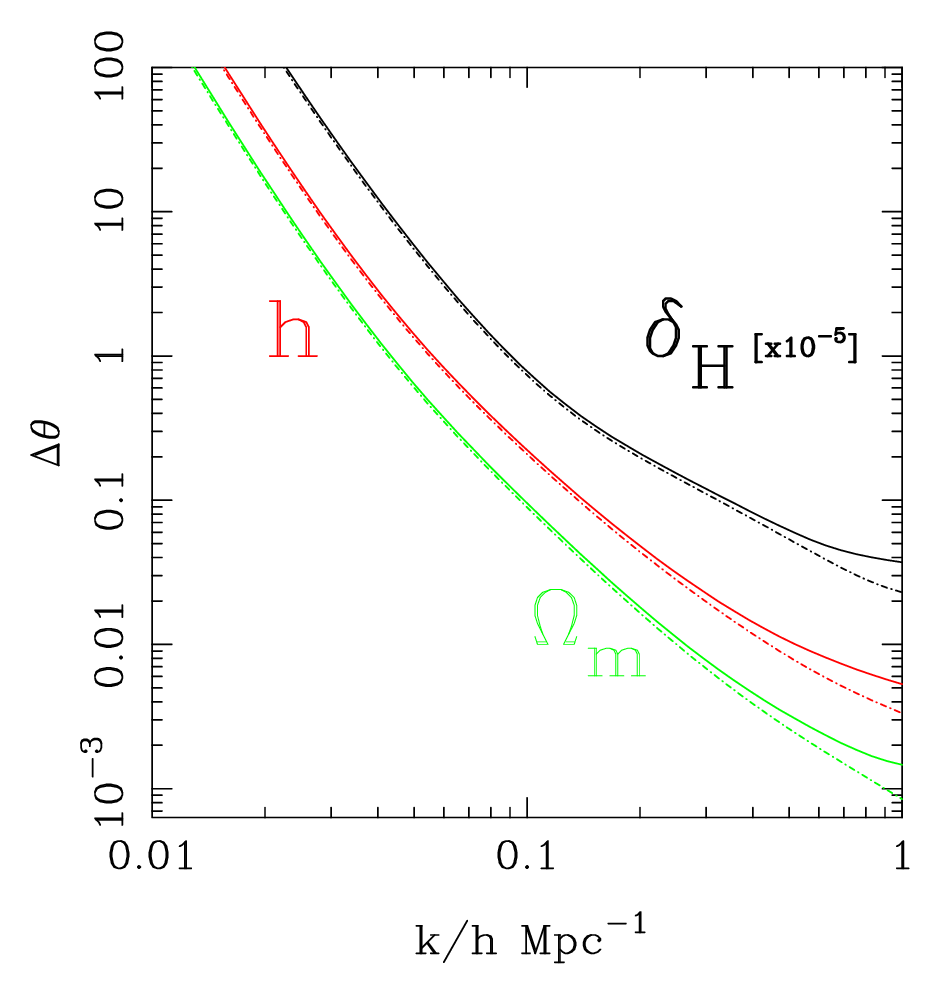,width=8.5cm,angle=0,clip=}}
\subfigure{\epsfig{figure=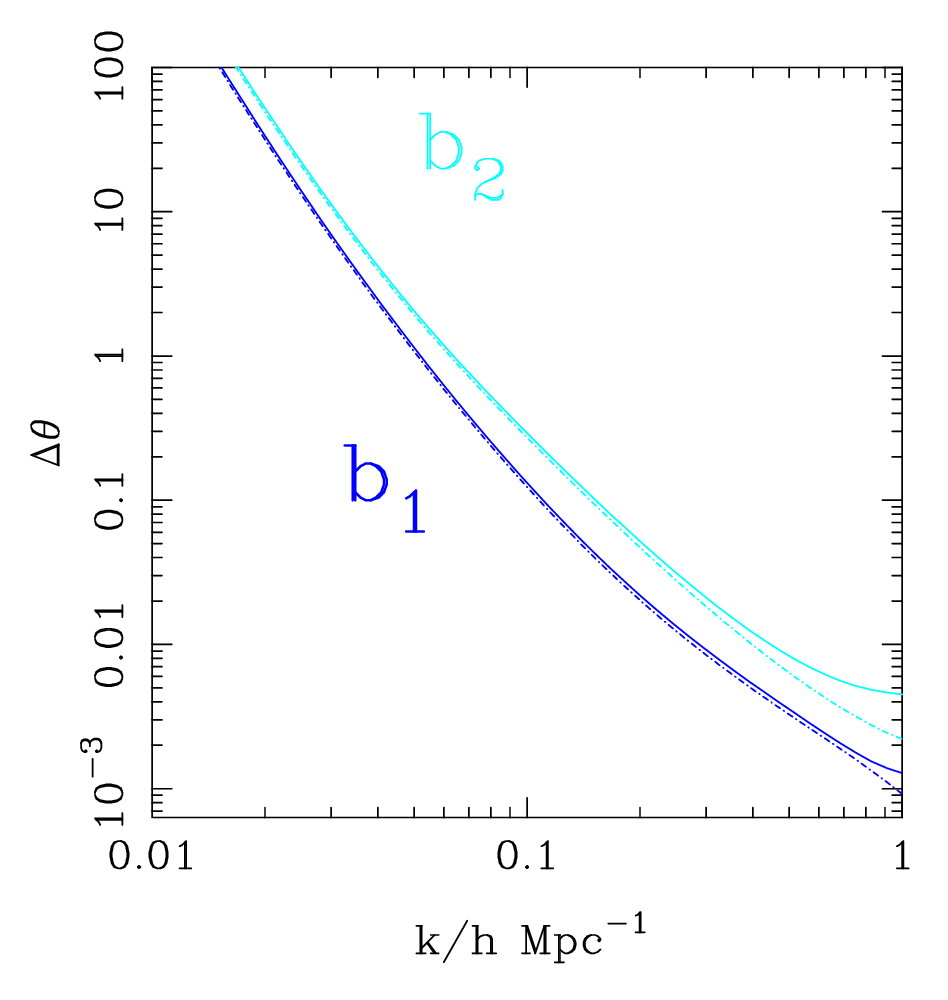,width=8.5cm,angle=0,clip=}}
\caption{Marginalised error estimates on 5 parameters obtained from a
Fisher analysis using the combined linear and nonlinear likelihood
with equilateral and degenerate triangles. The results are shown for two
sets of survey parameters resembling those of the 2dF (solid) and the
SDSS (dot--dashed). Table 1 contains the survey parameters.}
\label{fig4}
\end{figure*}

Marginal errors may be obtained from inversion of the Fisher matrix so 
that the the $1\sigma$ error on parameter $\theta_i$ is approximately given
by 
\be
\Delta \theta_i = \sqrt{(\F^{-1})_{ii}},
\label{errors}
\ee 
under the assumption that the likelihood distribution of
parameters is roughly Gaussian.  In Figure \ref{fig4} we show the
results of the Fisher analysis for the best attainable marginal
uncertainties on the 5 independent parameters
$\delta_H,h,\Omega_m,b_1$ and $b_2$ as a function of $k$.  Here, and
in subsequent figures, $k$ is the maximum mode, $k_{\rm max}$, in all
integrations.  A 5-D joint estimation of parameter uncertainties with
the Fisher matrix is only possible for a nonlinear likelihood since to
linear order the Fisher matrix is singular due to degeneracies between
parameters.
 
We give the results for two different redshift surveys which we define
by the volume of the survey, V, and the mean number density of
sources, $n$. Formally this represents a volume-limited survey, but
here we shall use effective values for a flux-limited sample. The
volume and number density are chosen to resemble those of the 2dF
redshift survey and the Sloan Digital Sky Survey. The relevant
parameters are found in Table \ref{table1}.

Interestingly, our results show that the accuracy of
the parameter estimation is similar for the 2dF and the SDSS, for all
of the parameters out to quite high k. The reason for this is that the
effective volumes of the two surveys are similar (see Table 1), and
the error scales like $1/\sqrt{V}$. Although the SDSS has a factor of
$\approx 3$ higher density of galaxies than the 2dF, this does not
become important until the models become noise dominated per
mode. From Figures 1 and 2, we see that for both surveys, this is not
until around $k=1\Mpch$, well below the scale where our parameter
estimation is valid.

Figure \ref{fig4} (LHS) shows the marginalised uncertainties on the
parameters $\delta_H$, $h$ and $\Omega_m$. At $k=0.1 \Mpch$, normally
the limit of linear analysis (Heavens \& Taylor 1995, Tadros et
al. 1999, Hamilton, Tegmark \& Padmanabhan 2000), the predicted
uncertainties lie around the $\approx 30\%$ level for both $h$ and
$\Omega_m$. Estimates for $\delta_H$, the horizon scale amplitude of
matter perturbations, fare a little better at around $20\%$. Extending
the analysis to $k=0.3 \Mpch$, near the limit of second-order
perturbation theory, we find the uncertainty falls by an order of
magnitude to $\approx 2\%$ for all three parameters. 

Figure \ref{fig4} (RHS) also shows the marginalised parameter
uncertainty for the bias parameters $b_1$ and $b_2$. For both surveys,
truncated at $k=0.1\Mpch$, we find that $b_2$ is not detected at all,
with $b_1$ detected at the $\approx 10\%$ level. However, if the
analysis is pushed to $k=0.3\Mpch$ the error on $b_1$ drops to around
a percent, while $b_2$ is now measured to around $5\%$
accuracy. While this looks encouraging for 2dF and the SDSS, we should
caution that we have assumed a volume limited survey, with
effective volume and noise levels chosen to mimic those of each
survey. A more accurate assessment of the improved accuracy will
require more detailed modeling and testing with N-body simulations,
although the present study should provide a good indication of the
measurement uncertainties.

%Figure \ref{fig4} (RHS) also shows the marginalised parameter
%uncertainty for the bias parameters $b_1$ and $b_2$. For a PSCz
%survey, truncated at $k=0.1\Mpch$, $b_2$ is not detected at all, and
%$b_1$ is only marginally detected with 2-$\sigma$ confidence. However,
%if the analysis is pushed to $k=0.3\Mpch$ the error on $b_1$ drops to
%around a few percent, while $b_2$ is now measured to around $10\%$
%accuracy. While this looks encouraging for the PSCz, we should caution
%that we have assumed a volume limited survey, with an effective volume
%and noise level chosen to mimic the PSCz. A more accurate assessment
%of the improved accuracy will require more detailed modeling and
%testing with N-body simulations, although the present study should
%provide a good indication of the measurement uncertainties.  The
%expectation for the accuracy of both bias parameters from the 2dF and
%SDSS is again around the few percent level for $k=0.3 \Mpch$, a drop
%of an order of magnitude from $k=0.1 \Mpch$, where the uncertainty on
%$b_1$ is of order $10\%$, and $b_2$ is again not detected.

\subsection{Lifting all the parameter degeneracies in redshift surveys}
\label{lifted}

\begin{figure}
\centering
\subfigure{\epsfig{figure=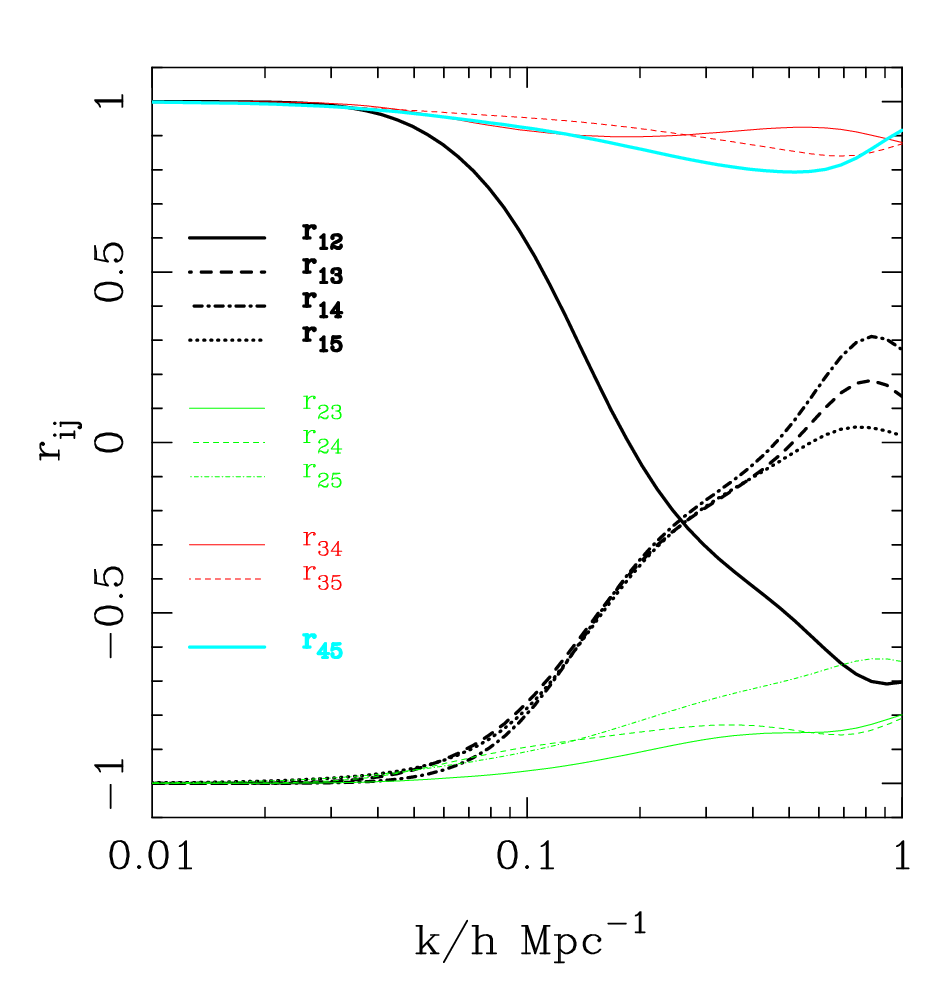,width=8.5cm,angle=0,clip=}}
\caption{Correlation coefficients for every pair of parameters in the
5-D marginalisation. The coefficients were calculated for the case of
a 2dF-type survey and for a combined linear and nonlinear likelihood
and both types of triangles. The key references \mbox{Table \ref{table2}} for
the index of each parameter.}
\label{fig5}
\end{figure}

As well as the marginal uncertainties it is interesting to consider 
the correlations between parameters. The information about such correlations 
lies in the off-diagonal components of the Fisher matrix. Defining the correlation
coefficient for parameters $\theta_i$ and $\theta_j$ as
\be 
	r_{ij} = \frac{(\eF^{-1})_{ij}}
	{\sqrt{(\eF^{-1})_{ii}(\eF^{-1})_{jj}}}, 
\ee
we can quantify the degree of correlation as a function of maximum
wavenumber. The results are plotted in Figure \ref{fig5} for all 10 of
the correlation coefficients, $r_{ij}$, in the 5-dimensional parameter
space considered above. The indices on the correlation coefficients
are given in Table \ref{table2}

At low-$k$ the Fisher matrix is dominated by the linear contribution
to the likelihood and subsequently all of the parameters remain
perfectly correlated/anti-correlated ($r_{ij} = \pm 1$). The reason
for this is that in the linear regime, one can only measure the three
composite parameters, $\beta = \Omega_m^{0.6}/b_1$, from
redshift-space distortions, $b_1 \delta_H$ from the amplitude of
linear galaxy clustering, and the spectral shape parameter,
$\Gamma=\Omega_m h$, which determines the break scale in the matter
power spectrum.

\begin{figure}
\centering
\subfigure{\epsfig{figure=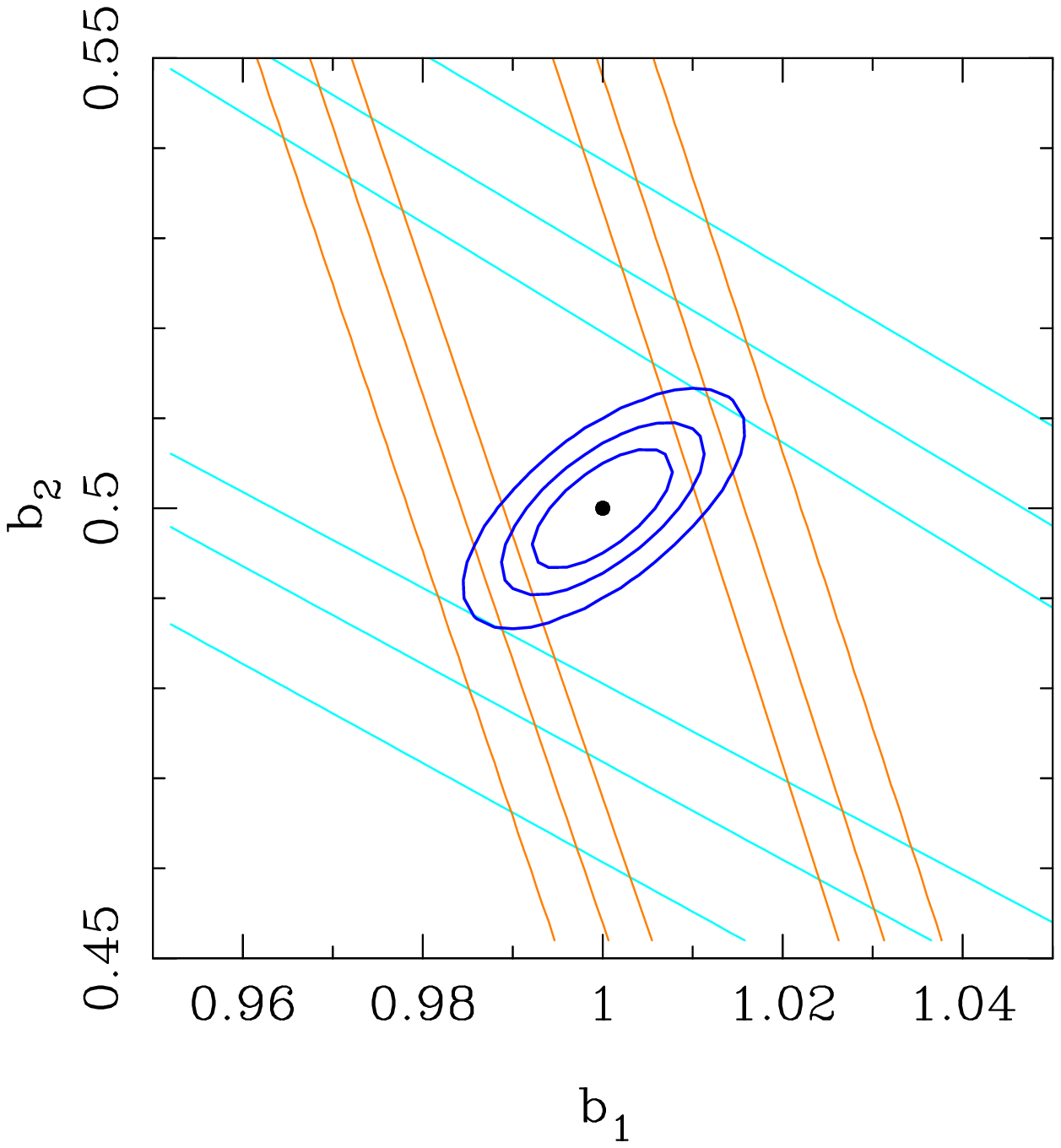,width=8.5cm,angle=0,clip=}}
\caption{Lifting the degeneracy  of the bias parameters using a combination
of different triangle shapes for the bispectrum. The contours denote
the 1, 2 and 3$\sigma$ confidence regions for Gaussian entropy plus
equilateral triangles and Gaussian plus degenerate triangles (faint lines) and
for the combined effect (dark lines). The entropy has been
marginalised over the amplitude $\delta_H$. Redshift distortions have
been ignored and  the noise and volume are similar to those of the 2dF}
\label{fig3}
\end{figure}

When the nonlinear terms begin to become important, at $k=0.04 \Mpch$,
the $r_{ij}$ deviate from $\pm 1$ indicating the decoupling of the
parameters.  The reason for this universal removal of degeneracies
lies in the lifting of the degeneracy between the bias parameters,
$b_1$ and $b_2$. With a single triangle shape, one can only measure
the combination $b_1/(1+b_2/2J(\k_1,\k_2)b_1)$ (Matarrese et al
1997). With a second triangle shape this constraint is lifted. Figure
\ref{fig3} illustrates this, where we have plotted the two degenerate
likelihood contours for $b_1$ and $b_2$ for equilateral and degenerate
triangles. The angle of intersection of these contours depends on the
choice of triangles and is maximised by choosing very different
shapes, motivating the choice of equilateral and degenerate
triangles. With the bias degeneracy removed, the degeneracies in
$\beta$, $\Gamma$ and $b_1 \delta_H$ are all lifted.

While the parameters are no longer degenerate, many remain quite
tightly correlated. The amplitude $\delta_H$ appears to become least
correlated, becoming almost completely decoupled from the other 4
parameters. This is because to linear order it is only degenerate with
$b_1$.

As $k$ gets larger, the $r_{ij}$ coefficients begin to show deviations
from a steady decline; in some cases this results in them actually
growing larger. This occurs because the correlations between
parameters ultimately depend upon shape and size of the 1-$\sigma$
likelihood surface around its maximum. As this surface shrinks with
increasing parameter information, it can change its shape, depending
on the relative response to clustering information. This is what
ultimately determines the correlation coefficient values.

\begin{table}
\begin{center}
\begin{tabular}{cc}\hline
 index & parameter \\ \hline 
1 & $\delta_H$ \\ 
2 & $h$ \\ 
3 & $\Omega_m$ \\
4 & $b_1$ \\ 
5 & $b_2$ \\ \hline
\end{tabular}
\caption{Cosmological parameters and their index for the correlation 
coefficient.}
\label{table2}
\end{center}
\end{table}

\subsection{A two-parameter model}
\label{twopara}
\begin{figure}
\centering
\subfigure{\epsfig{figure=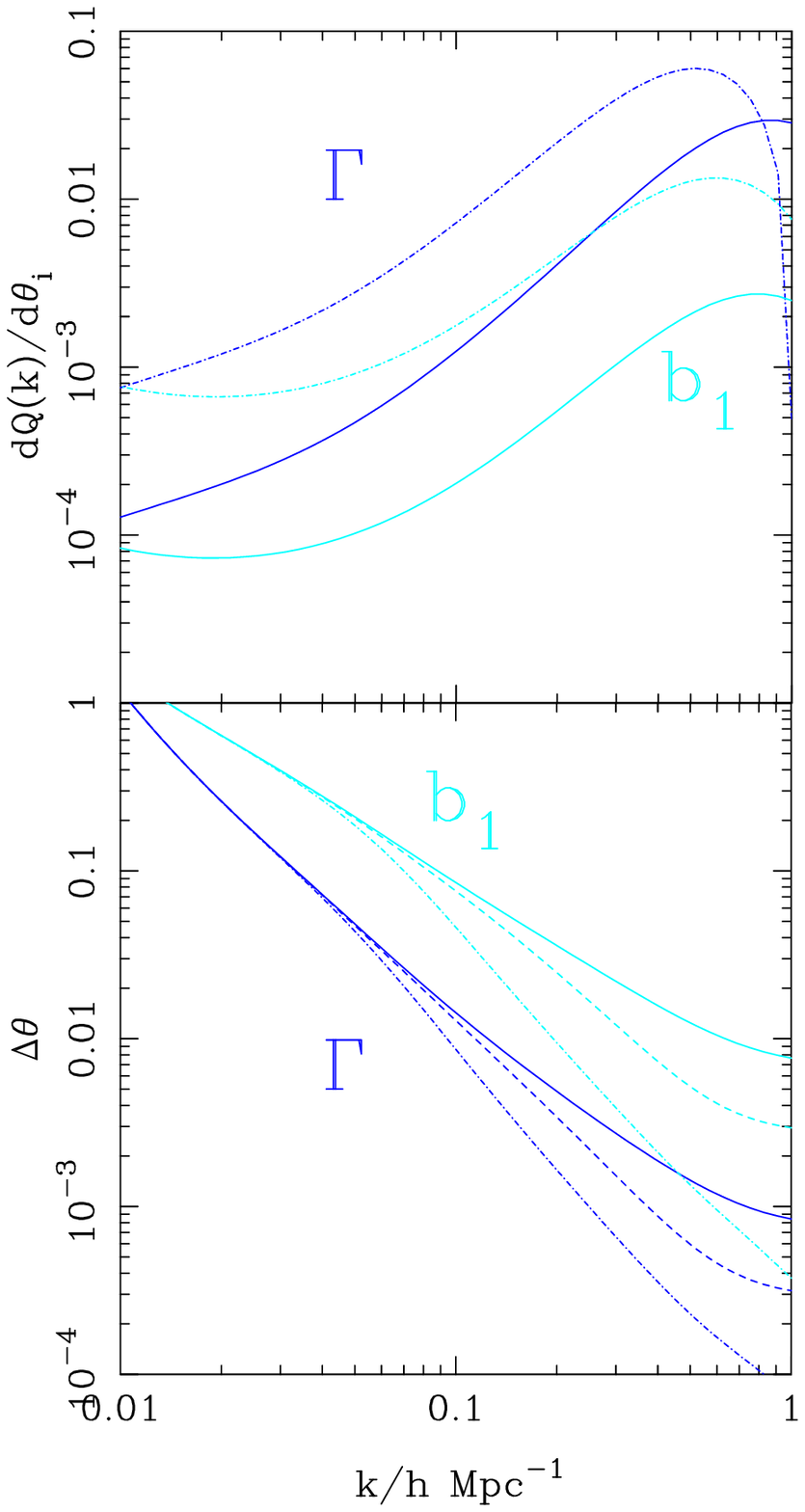,width=8.5cm,angle=0,clip=}}
\caption{Parameter derivatives (top panel) and 1$\sigma$ error levels
for the joint estimation of $b_1$ and $\Gamma$ in redshift space
(bottom panel).  The top panel shows the angle averaged derivative of
the $Q$ function from equation (\ref{Qdeff}) for equilateral triangles
(solid lines) and degenerate triangles (dotted lines) taken with
respect to each of the parameters.  The bottom panel shows the joint
errors on the parameters when using the Gaussian likelihood (solid
line), when adding in the contribution from equilateral triangles
(dashed line) and when adding in both equilateral and degenerate
triangles (dotted line).  The survey parameters closely resemble those
of the 2dF, as listed in Table \ref{table1}. }
\label{fig1}
\end{figure}

\begin{figure}
\centering
\subfigure{\epsfig{figure=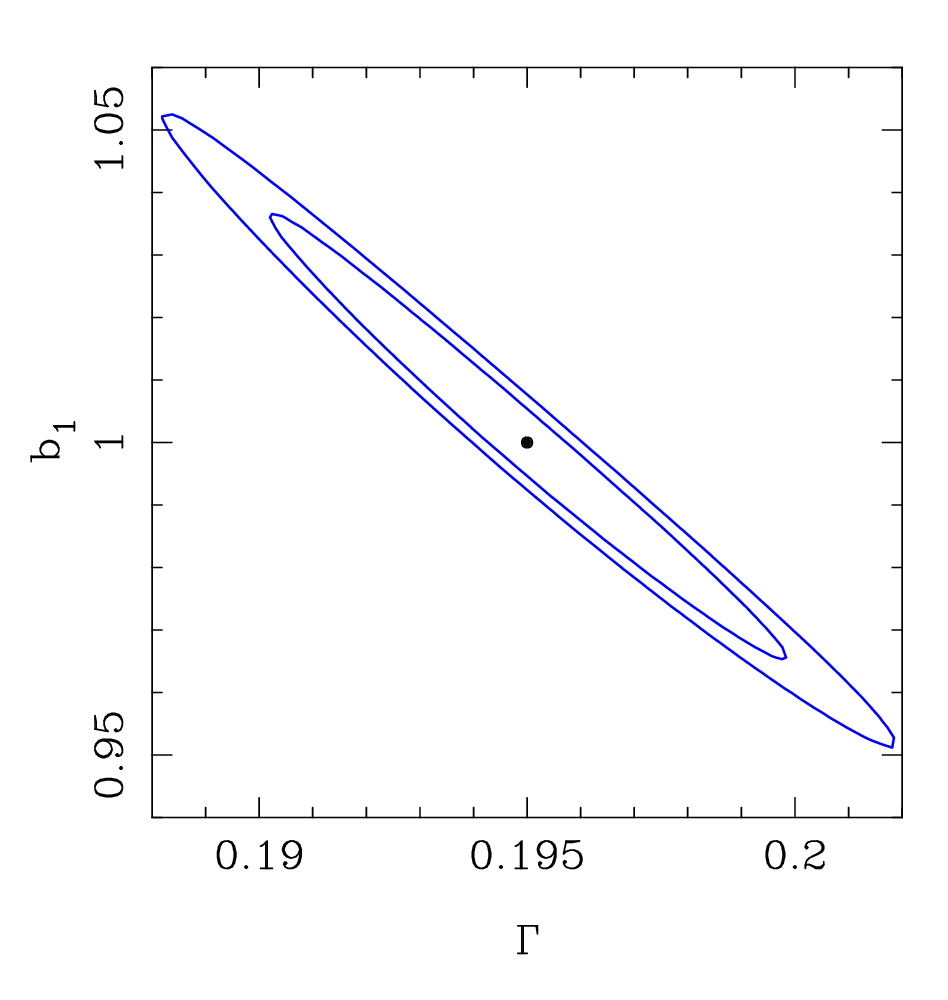,width=4.cm,angle=0,clip=}}
\subfigure{\epsfig{figure=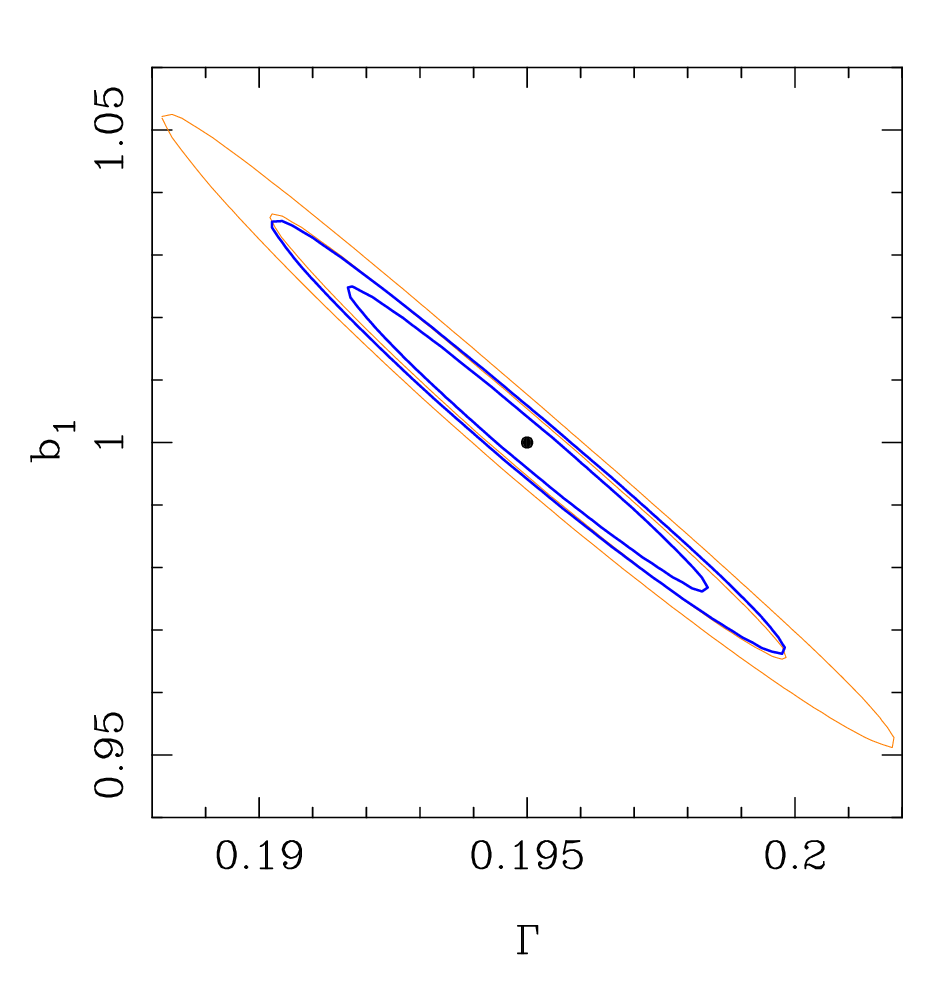,width=4.cm,angle=0,clip=}}
\subfigure{\epsfig{figure=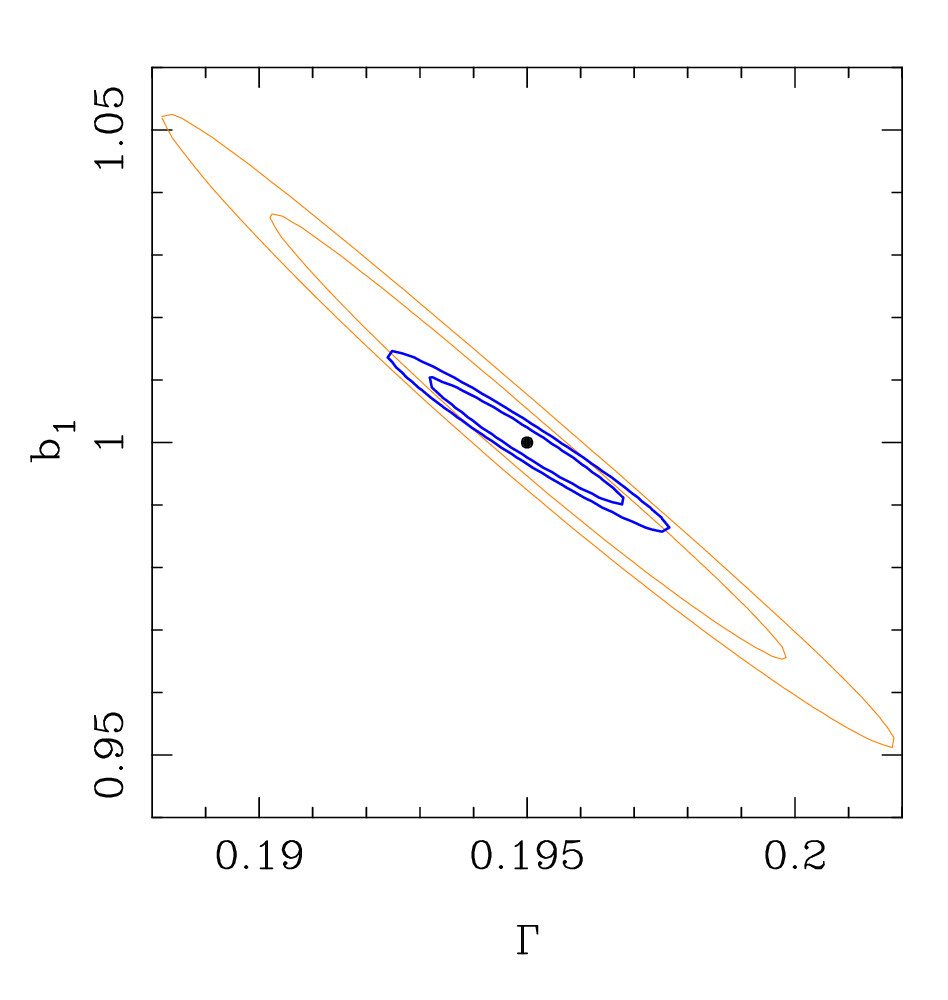,width=4.cm,angle=0,clip=}}
\subfigure{\epsfig{figure=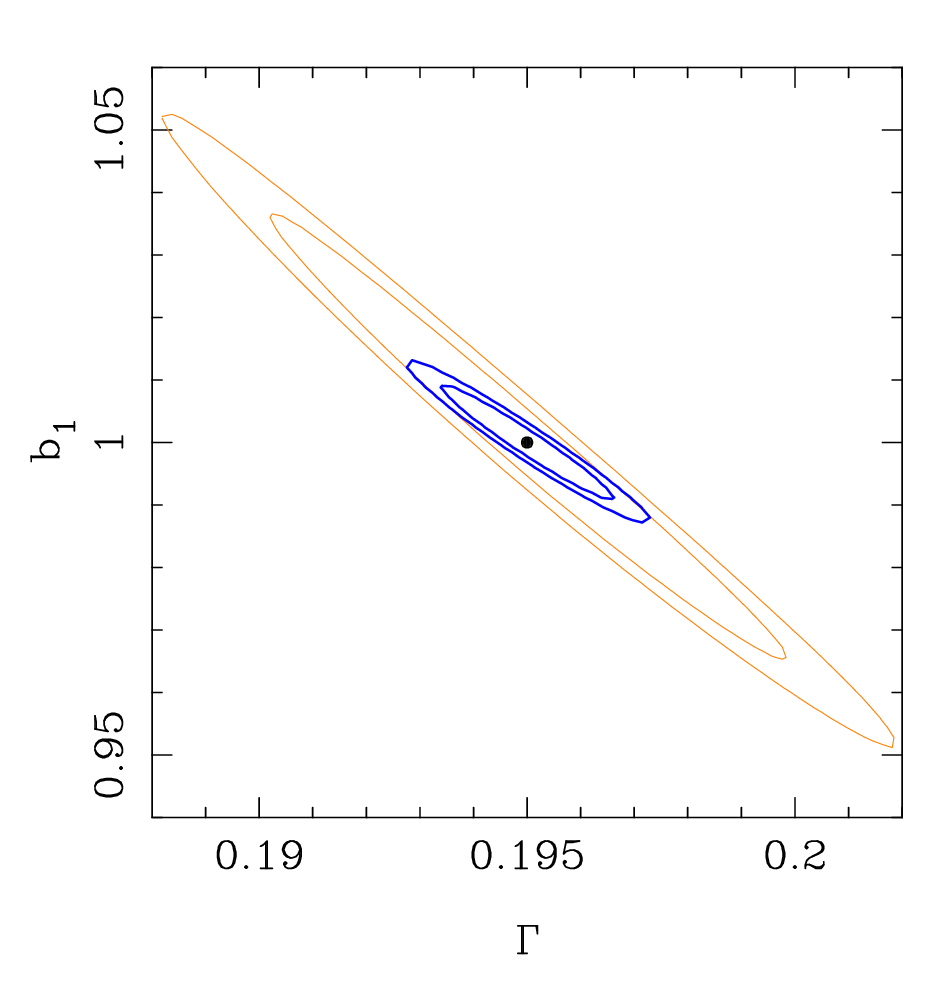,width=4.cm,angle=0,clip=}}
\caption{Entropy contours in the $b_1$ -- $\Gamma$ plane.  The contour
levels correspond approximately to 1 and 2$\sigma$ confidence
limits. From top left to bottom right the plots show: the Gaussian
entropy; the Gaussian entropy (light lines) and the contribution from
equilateral triangles (dark lines); the Gaussian entropy (light lines)
and the contribution from degenerate triangles; and the Gaussian
entropy (light lines) with the sum of the contribution from the
Gaussian part plus equilateral and degenerate triangles (dark
lines). For all plots $k_{max}$ was $0.2 \Mpch$. The survey parameters
closely resemble those of the 2dF. }
\label{fig2}
\end{figure}

While the Fisher matrix results show the value of including nonlinear
effects in parameter estimation, it is useful to examine in some more
detail how nonlinearity increases parameter information.  Rather than
deal with all of the parameters, we shall restrict our analysis to two
parameters. One is the linear bias factor, $b_1$, who's measurement
allows us to lift all the parameter degeneracies. The second is the
parameter combination, $\Gamma=\Omega_m h$. The advantage of
restricting our analysis to these two parameters is that their Fisher
matrix in the linear regime is not singular, and can be compared
directly with the nonlinear regime. We shall assume all other
parameters are know.

In Figure \ref{fig1} (top panel) we have plotted the derivatives of
the nonlinear, angle-averaged, redshifted function $Q(\k_1,\k_2)$,
given by equation (\ref{Qdeff}) with respect to each of the
parameters. This effectively shows the information content of the
modes, with respect to each parameter.  Dark lines show the derivative
with respect to $\Gamma$, while lighter ones with respect to
$b_1$. Solid lines are for equilateral triangles, while dotted lines
show the contribution from degenerate triangles.  Figure \ref{fig1}
shows that degenerate triangles have a higher derivative, indicating
that they contain more information that equilateral ones. We shall
return to this point in Section \ref{triinfo}. The similarity of the
shapes of the derivative also suggests that these parameters are
strongly correlated.

In Figure \ref{fig1} (bottom panel) we show the marginal $1\sigma$
uncertainties on $b_1$ and $\Gamma$ obtained from the Fisher analysis.
It is immediately striking that inclusion of just a single triangle
configuration has a significant effect on the errors. Even integrating
out to relatively modest wavenumbers, $k\approx 0.2\Mpch$, there may
be a 20 -- 30 \% reduction in the marginal uncertainty. This effect
continues to grow as more modes are included. Using both triangle
configurations has a much greater effect, reducing the errors by
nearly an order of magnitude out at $k = 0.3\Mpch$.

In Figure \ref{fig2} we plot the entropy contours for a slice in the
$b_1 - \Gamma$ plane, assuming the other parameters are known. The
top-left plot shows the Gaussian contribution to the parameter
entropy, where the maximum wavenumber is $k_{\rm max}=0.2 \Mpch$. The
two parameters are highly correlated, and have uncertainties of
$\Delta b_1 \approx 0.05$ and $\Delta \Gamma \approx 0.005$.  The
top-right plot show the effect of adding nonlinear information from
equilateral triangles; a $\approx 30\%$ improvement.  
%There is only a marginal improvement on the
%error ellipses.  
By adding only degenerate triangles to the Gaussian
term, in the bottom left figure, the uncertainty reduces more
dramatically. Finally, in the bottom-right plot we show the combined
effect of all of the terms.

Interestingly although the marginal errors are reduced, inclusion of
the nonlinear terms does not, in this case, make the two parameters,
$b_1$ and $\Gamma$, significantly less correlated, as suggested by the
parameter derivatives in Figure \ref{fig1}. This can also be seen from the
entropy contours in Figure \ref{fig2} which are all rather elongated.
%This is also suggested in the top panel of Figure \ref{fig1}. The
%similarity in shape of the curves suggests a significant correlation
%between parameters out to high $k$ for both equilateral and degenerate
%triangles. 
The origin of this degeneracy is the effect both parameters
have on the power spectrum. The linear bias factor, $b_1$, affects the
amplitude of the power spectrum (ignoring its effects on redshift
space distortions), while the shape parameter $\Gamma$ shifts the
break scale in wavenumber. If the power spectrum were a power-law,
these effects would be degenerate. But since there is a break, this
degeneracy is broken on scales below the break scale. Adding nonlinear
information, while reducing the absolute error, does not reduce this
correlation.

\subsection{The information content of triangles }
\label{triinfo}

It is apparent from Figures \ref{fig3} and \ref{fig1} that degenerate
triangles seem to contain a great deal more information about
cosmological parameters than equilaterals.  As this result seems
somewhat unintuitive, it is perhaps worth considering how the
bispectrum triangles add information to the Fisher matrix.

The effect of different triangles can be partly understood from
consideration of equation (\ref{nongausfish}). For the non-Gaussian
term, the Fisher matrix depends upon the derivatives of the function
$Q(\k_1,\k_2)$ weighted by the product of $P(k_1) P(k_2) P(|\k_1 +
\k_2|)$. For equilateral triangles this factor is the cube of the
power. However, for triangles where the magnitudes of the wavevectors
are not equal there is a mixing of power from different scales. This
implies is that information contained in a given triangle
configuration is related to the shape of the power spectrum. Although
the parameter derivatives of $Q(\k_1,\k_2)$ are larger for degenerate
triangles, indicating a larger information content (see Figure
\ref{fig1}), the power-weighting will ultimately determine which
contributes the most information.

For monotonically rising $P(k)$, equilateral triangles are most
significant because small scales (large $k$) contain most of the
power. Conversely for power spectra that are decreasing functions of
$k$, the maximum signal comes from large scales, hence for a given
$k_{max}$, triangles that have a small side will give the most weight
and provide the most information. In the CDM case the greatest power
lies around the break scale, $k_*$. The result is therefore an
admixture of the two competing effects; longwards of $k_*$ the
spectrum rises and equilateral triangles contain the most information,
shortwards and it is triangles that contain a short side that win
out. In the limit this means that for CDM there is a ``maximal''
triangle for parameter estimation that is degenerate, in the manner
discussed in Section (\ref{triangles}), with sides $k_{*},
k_{max},k_{max}-k_{*}$. In practise, estimating parameters from
degenerate triangles may be more difficult, as there will be fewer
independent low-$k$ modes within finite survey volume.

%In addition to the squeezing of parameter uncertainties, different
%triangle configurations appear to align the likelihood contours with
%different orientations in parameter space. This effect can be seen in
%figure \ref{fig2}. Compare the second and third panels which
%show the error ellipses for when just equilateral triangles (second
%panel) and just degenerate triangles (third panel) are used in
%conjunction with the linear entropy. This effect is useful when
%summing the contributions from more than one triangle configuration;
%the combined effect may lead to a further reduction in the marginal
%uncertainties and help in the breaking of degeneracies. 

\section{Conclusions}
\label{conclusions} 

We have presented a method for including higher-order moments in the
likelihood functions of cosmological fields in such a way that the
parameter dependencies of the non-Gaussian terms may be used for
estimating cosmological parameters. This non-Gaussian correction
generalises likelihood analysis for application to fields that either
contain intrinsic non-Gaussianity, or have become non-Gaussian due to
nonlinear gravitational evolution. Such a generalisation is
fundamentally important for cosmology since most of the interesting
fields show non-Gaussian properties. Our method has the advantage that
we attempt to deal with the probability distribution of the field
explicitly, allowing the natural combination of linear and nonlinear
regimes. In this respect our work differs significantly from existing
techniques that separate out the nonlinear regime and apply a Gaussian
PDF to the higher-order moments themselves (Matarrese et al. 1997,
Verde et al. 1998, Scoccimarro et al. 1999, Scoccimarro et al. 2000).

We have presented a general formalism for calculating the non-Gaussian 
Fisher matrix and the shape of the likelihood around its maximum, the 
parameter entropy function. A central result of this paper has been 
to show that to lowest order, the dominant effect of
adding nonlinearity, or more generally a non-Gaussianity, is to increase the
the parameter information. This effect dominates over the degradation
of parameter information due to the nonlinear evolution of the shape of the
likelihood function.

Applying our analysis of nonlinearity to a simple model for a galaxy
redshift survey, including the effects of shot-noise, redshift-space
distortions and galaxy bias, we have found that nonlinear effects may
be extremely useful for placing tight constraints upon cosmological
parameters.
% and should improve parameter estimation by an order of
%magnitude. 
Of crucial importance is the fact that even at second-order,
degeneracies can be broken so that all of the cosmologically
interesting parameters may be estimated independently. While going to
higher-order may place tighter constraints on the nonlinear bias
function through estimation of the series coefficients $b_n$,
essentially redshift surveys offer up a great deal of their
information without the need to turn to progressively more intricate
perturbative calculations. In addition, this analysis suggests that
galaxy redshift surveys can be used in isolation to determine
cosmological parameters. This greatly enhances their power as
parameter estimation can be compared from independent surveys, such as
the CMB, cosmic velocity fields and weak lensing surveys and combined
further to reduce uncertainties.

Our analysis relies upon a multivariate generalisation of the
Edgeworth approximation for the probability distribution function.  As
such one should be careful to determine the range of applicability of
second-order perturbation theory, and the series expansion of the
likelihood function.  It is well known from studies of 1-point
distributions (Bernardeau \& Kofman 1995, Juszkiewicz et al. 1995
Gazta\~{n}aga et al. 1999, Taylor \& Watts 2000, Watts \& Taylor 2000)
that the Edgeworth PDF has limitations -- displaying unphysical
features when the variance or skewness becomes too high. However,
these effects do not automatically render the Edgeworth a bad
approximation. Comparison with N-body simulations in redshift-space
(Watts \& Taylor 2000) has shown that peak of the distribution remains
a good fit even when the extremities begin to behave
badly. Fortunately we have found that a good deal of information comes
from the nonlinear terms at relatively low-$k$ and have shown that the
degeneracies between parameters are broken even for modest values of
wavenumber. This is good news from a computational point of view as
well since the number of modes available for analysis grows like
$k^3$. While this is beneficial in some respects, it can also lead to
computational problems if the number of modes gets too high. In his
case data compression methods may need to be incorporated (Taylor et
al. 2000).  In a forthcoming paper (Watts \& Taylor, in preparation)
we rigorously test our model on N-body simulations in order to
accurately determine the limitations, and address the specific issues
associated with application to real data sets.

\section*{Acknowledgements}
PIRW thanks the PPARC for a research studentship. ANT is a PPARC
Advanced Fellow. The authors thank Alan Heavens for
useful discussion about the bispectrum.

\section*{References}

\bib Amendola L., 1996, MNRAS, 283, 983

\bib Bardeen J.M., Bond J.R., Kaiser N., Szalay A.S., 1986, ApJ, 304, 15

\bib Bernardeau F., Kofman L., 1995, ApJ, 443, 479

\bib Bond J.R., Efstathiou G., Tegmark M., 1997, MNRAS, 291, L33

\bib Bouchet F.R., Juszkeiwicz R., Colombi S., Pellat., 1992, ApJ, 394, L5

\bib Coles P., Jones B., 1991, MNRAS, 248, 1

\bib Colless M., 1998, In proc. 14th IAP meeting, Paris

\bib Dekel A., Lahav O., 1999, ApJ, 520, 24

\bib Folkes S., Ronen S., Price I.,
        Lahav O., Colless M., Maddox S., Deeley K.,
        Glazebrook K., Bland-Hawthorn J., Cannon R., 
	Cole S., Collins C., Couch W., Driver S.P., 
	Dalton G., Efstathiou G.,
        Ellis R.S., Frenk C.S., Kaiser N.,
        Lewis I., Lumsden S., Peacock J.A.,
        Peterson B.A., Sutherland W., Taylor K., 1999, MNRAS, 308, 459

\bib Fry J.N., 1994, Phys Rev Lett, 73, 215

\bib Fry J.N., Gazta\~{n}aga E., 1993, ApJ, 413, 447

\bib Gazta\~{n}aga E., Fosalba P., Elizalde E., 2000, ApJ, 539, 522

\bib Goroff M., Grinstein B., Rey S., Wise M.B., 1986, ApJ, 311, 6

\bib Gunn J.E., 1995, In proc. American Astronomical Society Meeting, 
	186, 4405 

\bib Hamilton A.J, Tegmark M., Padmanabhan N., 2000, MNRAS, 317, L23 

\bib Heavens A.F., Taylor A.N., 1995,  MNRAS, 275, 483

\bib Heavens A.F., Taylor A.N., 1997, MNRAS, 290, 456

\bib Heavens A.F., Matarrese S., Verde L., 1998,  MNRAS, 301, 797

\bib Hivon E., Bouchet F.R., Colombi S., Juszkiewicz R., 1995, ApJ, 298, 643

\bib Hu W., Tegmark M., 1999, ApJ, 514, L65

\bib Jungman G., Kamionkowski M., Kosowsky A., Spergel D.N., 1996, Phys Rev D,
	54, 1332

\bib Juszkeiwicz R., Weinberg D.H., Amsterdamski P., 
	Chodorowski M., Bouchet F., 1995, ApJ, 442, 39

\bib Kaiser N., 1987, MNRAS, 227, 1

\bib Kendal M.G., Stuart A., 1969, ``The Advanced Theory of Statistics, 
	Vol. 2'', Griffin, London

\bib Ma C., Fry J.N., 2000, astro-ph/0003343

\bib Maddox, S., 1998, In proc. ASP Conf. Ser. 146: The Young Universe: 
	Galaxy Formation and Evolution at Intermediate and High Redshift

\bib Matarrese S., Verde L., Heavens A.F., 1997, MNRAS, 290, 651

\bib Peacock J.A., 1992, MNRAS, 258, 581

\bib Peacock J.A., Dodds S.J., 1994, MNRAS, 267, 1020

\bib Peacock J.A., Smith R.E., 2000, astro-ph/0005010

\bib Peebles P.J., 1980, ``Large-Scale Structure in the Universe'',
   Princeton University Press, Princeton

\bib Pen U., 1998, ApJ, 504, 651

\bib Rocha G., Magueijo J., Hobson M., Lasenby A., 2000, astro-ph/0008070

%\bib Saunders W., Sutherland W.J., Maddox S.J., 
%        Keeble O., Oliver S.J., Rowan-Robinson M., 
%        McMahon R.G., Efstathiou G.P., Tadros H., 
%        White S.D.M., and Frenk C.S., Carrami{\~n}ana A.,
%        Hawkins M.R.S., 2000, MNRAS, 317, 55

\bib Scoccimarro R., Colombi S., Fry J.N.,
        Frieman J.A., Hivon E., Melott A., 1998, ApJ, 496, 586

\bib Scoccimarro R., Couchman H.M.P., Frieman J., 1999, ApJ, 517, 531

\bib Scoccimarro R., Feldman H.A., Fry J.N., Frieman J., 2000 (astro-ph/0004087)

\bib Seljak U., 2000, astro-ph/0001493

\bib Tadros H., Ballinger W.E., Taylor A.N., Heavens A.F., Efstathiou G.,
    	Saunders W., Frenk C.S., Keeble O., McMahon R., Maddox S.J., 
        Oliver S., Rowan-Robinson M., Sutherland W.J., 
        White S.D.M., 1999, MNRAS, 305, 527

\bib Taylor A.N., Watts P.I.R., 2000, MNRAS, 314, 92 

\bib Taylor A.N., Ballinger W.E., Heavens A.F.,Tadros H., 2000, 
	astro-ph/0007048

\bib Tegmark M., 1997, Phys Rev Lett, 79, 3806

\bib Tegmark M., Taylor A.N., Heavens A.F., 1997, MNRAS, 480, 22

\bib Verde L., Heavens A.F., Matarrese S., Moscardini L., 1998, MNRAS, 300, 747

\bib Watts P.I.R., Taylor A.N., 2000, astro-ph/0006192

\onecolumn

\section*{Appendix A}
For completeness, we provide the general continuum form for the
non-Gaussian perturbation term appearing in equation (\ref{perturb}) 
\ba
	X & = & \int\! \prod_{i=1}^4 d^3 \! x_i \, D(\x_1,\x_2,\x_3)\,
	C^{-1}(\x_4,\x_1)  C^{-1}(\x_2,\x_3)\,\phi(\x_4) + 
	{\rm cyc}(231,312) \nn
 	& + &
	\int\! \prod_{i=1}^6 d^3 \! x_i\,  D(\x_1,\x_2,\x_3)\, 
	C^{-1}(\x_1,\x_4)C^{-1}(\x_2,\x_5)
	C^{-1}(\x_3,\x_6)\,\phi(\x_4)\phi(\x_5)\phi(\x_6)
\label{Xfact}
\ea
where ${\rm cyc}$ means to permutate indices.

\section*{Appendix B}
In the absence of noise and redshift distortions the nonlinear Fisher
matrix and entropy can be written in a particularly pleasing form. The
Fisher matrix for equilateral triangles reduces to
\be
	\eF_{ij} =  \frac{V}{2}\!\int \!\!\frac{k^2 d k}{2 \pi^2}
	\Big[1+\Big(\frac{12}{7}\Big)^2 \frac{\Delta^2(k)}{6}\Big]\, 
 	[\de_i \ln{\Delta^2(k)}][\de_j \ln{\Delta^2(k)}],
\ee
while the parameter entropy is given by
\be
 	S = \frac{V}{2} \int \! \frac{k^2 d k}{2 \pi^2}\, 
	\Big\{ \ln{\Delta^2(k)}+\frac{\Delta'^{2}(k)}{\Delta^2(k)}
 	 + \frac{1}{12}\Big(\frac{12}{7}\Big)^2\Delta'^{2}(k)\Big[
	\Big(\frac{\Delta'^{2}(k)}{\Delta^2(k)}\Big)^2 - 
	\frac{2 \Delta'^2(k)}{\Delta^2(k)}\Big]\Big\}.
\ee
For the case of degenerate triangles the Fisher matrix becomes
\ba
\eF_{ij} & = &  \frac{V}{2}\!\int \!\!\frac{k^2 d k}{2 \pi^2}\Big\{
	\Big[1+\frac{8}{3} \frac{\left(\Delta^2(k/2)\right)^2}
	{\Delta^2(k)}\Big]\; 
 	[\de_i \ln{\Delta^2(k)}][ \de_j \ln{\Delta^2(k)}]
	+ \frac{1}{6}\Delta^2(k)\; 
	[\de_i \ln{\Delta^2(k/2)}][ \de_j \ln{\Delta^2(k/2)}] \nn 
	& & \; \; \; - \frac{32}{3}\Delta^2(k/2)\;  
	\Big(\de_{[i} \ln{\Delta^2(k/2)}\; \de_{j]} \ln{\Delta^2(k/2)}\Big) \Big\},
\ea
and the parameter entropy is 
\ba
S & = &  \frac{V}{2} \int \! \frac{k^2 d k}{2 \pi^2}\, \Big\{
\ln{\Delta^2(k)}+\frac{\Delta'^{2}(k)}{\Delta^2(k)} + 
\frac{1}{12}\Delta'^2(k/2) \Big[\frac{\Delta'^2(k/2)\Delta'^2(k)}
{[\Delta^2(k)]^2} - \frac{1}{2}\frac{\Delta'^2(k/2)\Delta'^2(k)}
{\Delta^2(k/2)\Delta^2(k)} + \frac{1}{16}\frac{\Delta'^2(k/2)\Delta'^2(k)}
{[\Delta^2(k/2)]^2}  \nn
 & & \, \, -  
\frac{1}{2}\frac{\Delta'^2(k/2)}{\Delta^2(k)}  -  
\frac{1}{32}\frac{\Delta'^2(k)}{\Delta^2(k/2)} +
\frac{1}{8}\frac{\Delta'^2(k/2)}{\Delta^2(k/2)} +
\frac{1}{8}\frac{\Delta'^2(k)}{\Delta^2(k)} \Big] \Big\}.
\ea

\section*{Appendix C: Redshifted Bispectrum}	

Here we present some of the formulae used to calculate the 
redshift space corrections for the bispectrum. 
The kernel from equation (\ref{bisp_eek})  for equilateral triangles is,
\ba
	{\rm Ker} &=& 3 \, \Big\{\frac{2}{7}  + \frac{1}{4}\beta (9/7+b_1)  
	+ \frac{3}{112}  \beta^2 \Big[ 3-7 b_1 (2-b_1)\Big] 
	 + \frac{1}{224} \beta^3 \Big[\mu^2(3-4\mu^2)^2 + 7 b_1 (1-b_1^2)
	(9(1-\mu^2) + 8 \mu^4 (3-2 \mu^2))\Big] \nn
	& &	
	- \frac{1}{256}\beta^4 (b_1-1)  (27 - 144 \mu^2 + 384 \mu^4 - 256 
	\mu^6) \Big\}.
\ea
The nonlinear small scale damping factor is
\ba
	D_2(k \sigma \mu) &=&\frac{1}{\sqrt{1+k^2 \sigma^2 \mu_1^2/2}}
  	\frac{1}{\sqrt{1+k^2 \sigma^2 \mu_2^2/2}}
	\frac{1}{\sqrt{1+k^2 \sigma^2 \mu_3^2/2}}\nn
	&=&
	\frac{1}{\sqrt{1+3/4 k^2 \sigma^2 + 
	9/64 k^4 \sigma^4 + k^6 \sigma^6 \mu^2(3-4\mu^2)^2}}
\ea

The calculation in Section \ref{redshift} can be repeated for
degenerate triangles. In this case the angles between the line of
sight and the wavevectors of the bispectrum triangles are 
\ba	
	\mu_1 &=& \mu, \nn
	\mu_2 &=& -\mu \nn
	\mu_3 &=& -\mu
\ea
The redshifted bispectrum is then given by
\ba
	B^s_g(\k) &=& D_2(k,\mu)\Big[2 b_1^3 {\rm Ker}
	(k,\mu,\beta,b_1) 
	+ b_2 b_1^2(1+ \beta \mu^2)^2(2P(k) +P(k/2))\Big]P(k/2) \nn
	& & + \frac{1}{n} b_1^2 D_1(k,\mu_1) (1+\beta\mu_1^2)^2 P(k) 
       +\frac{1}{n} b_1^2 D_1(k,\mu_2) (1+\beta\mu_2^2)^2 P(k/2) 
       +\frac{1}{n} b_1^2 D_1(k,\mu_3) (1+\beta\mu_3^2)^2 P(k/2) 
  + \frac{1}{n^2},
\label{redshiftedbideg}
\ea
where in this case the kernel is
\ba
	{\rm Ker} &=& \frac{1}{2}\Big[4 P(k/2) - P(k)\Big] + \frac{1}{2}\mu^2
	\Big[ 4(3+b_1)P(k/2) - 3 P(k)\Big] \beta + \frac{1}{2}\mu^4\Big[
	(4 b_1-4 b_1^2 -3) P(k) + 2(5b_1 + b_1^2 + 6)P(k/2)\Big]\beta^2 \nn
	& & - \frac{1}{2}\mu^6\Big[(1-8b_1+8b_1^2) P(k) - 
	4(1+b_1^2)P(k/2)\Big]\beta^3 - 
	b_1 \mu^8\Big[ (2 - 2 b_1)P(k) +  (1+b_1) P(k/2)\Big]\beta^4,
\ea
and the damping term
\be
	D_2(k,\mu) = \frac{1}{\sqrt{(1+k^2 \sigma^2 \mu^2/8)(1+k^2\sigma^2
	 \mu^2/2)}}
\ee

\end{document}